\documentclass[%
 reprint,
 amsmath,amssymb,
 aps,prb
]{revtex4-1}
\usepackage{graphicx}
\usepackage{dcolumn}
\usepackage{bm}

\usepackage{xcolor}
\begin{document}
\title{Anomalous Josephson effect and quantum anomaly in inversion asymmetric Weyl semimetals}
\author{Debabrata Sinha}
\affiliation{Center for Theoretical Studies, Indian Institute of Technology,
Kharagpur-721302, India}
\date{\today}
\begin{abstract}
We study a Josephson junction involving an inversion-asymmetric Weyl semimetal in presence of time-reversal symmetric (TRS) or time-reversal symmetry broken tilt in the Weyl spectra. We reveal that both types of tilts in the Weyl nodes lead to a Josephson $0$-$\pi$ transition and a zero bias valley/chiral supercurrent. Strikingly, the TRS tilt gives rise to a pure valley Josephson current (VJC) and TRS broken tilt induces a pure chirality Josephson current (CJC) in this system. The VJC and CJC are the manifestation of valley symmetry broken and $\mathbb{Z}_2$ symmetry broken by the respective tilt. We obtain the reversal of a pure VJC and pure CJC even in the zero bias condition controllable by the junction length.  Our analysis of controllability of valley and chirality dependent transport in an inversion asymmetric Weyl semimetal junction could allow applications in valleytronics and chiralitytronics, respectively. The tilt induced Josephson effect provides an alternative route for supercurrent $0$-$\pi$ transition, different from the conventional ferromagnetism Josephson junctions where the spin polarization is essential. In the long junction and zero temperature limit, VJC and CJC are associated with a quantum anomaly which is manifested through a discontinuous jump in the current in absence of TRS and TRS breaking tilts, respectively.
\end{abstract}
\maketitle

\textbf{Introduction:-} The recent discovery of Weyl fermions in a number of materials draw intensive attention owing to their novel physics associated with Weyl nodes\cite{Huang-NatCom15,Xu-Sience15,Weng-PRX15,Yang-NatPhy15,Hirayama-PRL15,Ruan-Nat16,Burkov-PRL11,Wan-PRB11,Xu-Science15,Lv-PRX15}. Weyl semimetals (WSMs) are gapless topological materials whose low energy excitations are Weyl fermions, particles that play significant roles in quantum field theory. The energy spectra show a pair of strongly degenerate Weyl cones having opposite chirality separated in momentum space. The stability of Weyl nodes requires at least time-reversal (TR) or inversion (IR) symmetry is broken in the system. The minimal model of TR broken WSM contains a single pair of Weyl nodes, whereas, an IR broken WSM includes four Weyl nodes with total zero chirality\cite{Armitage-RMP18,Balents-PRB12}. Chirality is an intrinsic property of Weyl fermions and can be understood as the topological charge of a Weyl node. The possibility to probe and manipulate the chirality and valley of a WSM thus remains an important issue in this context. Several studies have been devoted to such chirality and valley-dependent physics of WSMs, recently\cite{Ma-Nat17,Ghosh-PRB20,Trauzettel-PRL18,Yang-PRL15,Heidari-PRB20,Simon-PRB19,Simon-PRB17}.

The topological nontrivial nature provides promising transport properties of WSMs including anomalous hall effect\cite{Pesin-PRL17,Burkov-PRL14}, Andreev reflection\cite{Uchida-Jpn14,Bovenzi-PRL17}, magnetotransport\cite{Zyuzin-PRB12,Son-PRB13}. Josephson junction presents another complementary route to investigate the anomalous transport properties of topological superconductors. Josephson junctions based on Dirac/Weyl semimetals have been investigated\cite{Yu-PRL18,Li-Nat18,Madsen-PRB17,Khanna-PRB16}, recently. To date, most of the experimentally discovered WSMs are inversion-asymmetric which includes TaAs class of materials\cite{Huang-NatCom15,Xu-Sience15,Yang-NatPhy15,Hirayama-PRL15,Weng-PRX15,Ruan-Nat16}. The chirality of Cooper pairs is a well-defined property in a Josephson junction of an inversion-asymmetric WSM since an s-wave superconductor connects the Weyl nodes of the same chirality\cite{Trauzettel-PRL18}. This is different from the TR-broken Weyl semimetals where BCS-pairing couples two Weyl nodes of opposite chirality\cite{Sinha-PRB20}.

The linear energy dispersion near Weyl nodes generally tilted along a certain momentum direction. The Lorentz symmetry is spontaneously broken in WSMs by the tilted dispersion\cite{Soluyanov-Nat15,Jiang-Nat17,Wang-PRL16,Li-Nat17}. Weyl fermions are categorized into two types, type-I and type-II, depending on whether the tilts exceed the Fermi velocity of electrons or not. Although the tilts of Weyl cone does not change the topology of the energy bands, it largely influences the quantum transports\cite{Hou-PRB17,Faraei-PRB19,Faraei-PRB20,Sinha-PRB20,Sinha-EPJB,Chan-PRB17,Beenakker-PRL16,Trescher-PRB15}. One of the intriguing phenomena occurs in a Josephson junction of TR broken WSM, where Cooper pairs acquire an extra momentum due to tilt\cite{Sinha-PRB20}. This brings an unusual oscillation in the Josephson current including Josephson's $0$-$\pi$ transition.

Thus it is natural to ask whether tilt can lead to anomalous effects in the Josephson junction involving inversion-asymmetric WSMs. Particularly, whether tilts can probe the chirality and valley physics of WSMs. The chirality Josephson effect of an inversion-asymmetric WSM has been studied in Ref.\cite{Trauzettel-PRL18}, by introducing Zeeman term. The gauge field due to the Zeeman term couples anti-symmetrically to the spin. This causes a finite chirality of Josephson current in the system. Here, we show that tilt causes the chirality as well as the valley dependent Josephson effects by shifting the Weyl nodes. No magnetic/Zeeman term is required to explain this shifting.

In this article, we study the Josephson effect of an inversion broken WSM Josephson junction with proximity induced s-wave superconductor. The interplay between the tilts, valley, and chirality of Weyl nodes, and s-wave superconductor resulting in unusual behaviors of a supercurrent through the junction. We find that tilt introduces an extra phase in the current-phase relations (CPRs). The tilt induced phase causes supercurrent reversal and Josephson $\phi$ junction in each valley and chirality sectors. The phase shift is controllable by the junction length and doping, for a finite value of tilt. The current phase relation is dominated by second harmonics in the supercurrent $0$-$\pi$ transition. The tilt-induced phase may occur in absence of any magnetic term. Therefore, these anomalous Josephson effects can exist in time-reversal set up also. Furthermore, the tilt induced phase shifts the current phase relations (CPRs) of opposite valleys and chirality differently. The unequal phase initiates many exotic phenomena. A finite VJC and CJC develop, respectively, in presence of TRS and TRS broken tilts. A pure VJC and CJC reversal are possible by tuning the junction length, even in zero bias condition. In the long junction and zero temperature limit, a singularity of VJC and CJC when the tilt induced phase disappears signifies a quantum anomaly of Cooper pairs.

\textbf{Model Hamiltonian and setup:-} Consider the general low-energy Hamiltonian describing a single Weyl node,
\begin{eqnarray}
H_{W}(\mathbf{q})=\hbar v_tk_t\sigma_0+\hbar v_F\hat{v}_{i,j}k_i\sigma_j
\label{gen-hamil}
\end{eqnarray}
where $v_F$ is the Fermi velocity without tilt. $\sigma_0$and $\sigma_j=\sigma_1,\sigma_2,\sigma_3$ are the $2\times 2$ identity matrix and Pauli matrices, respectively. $\hat{v}_{i,j}$ is the anisotropy in the spectrum and $\chi=Det(\hat{v}_{i,j})=\pm 1$ determines the chirality of the given node. The quasiparticle momentum components are $k_i$ and the momentum along tilt direction is $k_t$. The energy dispersion of Eq.(\ref{gen-hamil}) is given by $E_{\pm}(\mathbf{k})=\hbar v_tk_t\pm \hbar v_F\sqrt{\sum_{i,j}k_i(\hat{v}\hat{v}^T)_{ij}k_j}=T(\mathbf{k})\pm U(\mathbf{k})$, where $T(\mathbf{k})$ and $U(\mathbf{k})$ regarded as the kinetic and potential parts. The ratio $v_t/v_F$ measures the tilt of the Weyl cone. The Weyl cones are classified in type-I if $T(\mathbf{k})< U(\mathbf{k})$ (i.e., $v_t< v_F$) and in type-II if $T(\mathbf{k})> U(\mathbf{k})$(i.e., $v_t> v_F$)\cite{Soluyanov-Nat15}. Here, we restrict our study for type-I WSMs.



An inversion broken but TR preserved WSM has a minimum of four Weyl nodes. The TR pair of Weyl nodes have the same chirality and if one of them appears at $\mathbf{q}$ then another must be at $-\mathbf{q}$. The total chirality is zero guarantees that there must be another TR pairs of Weyl nodes with opposite chirality. The mirror symmetry relates the opposite chiral Weyl nodes by flipping the sign of momentum along the mirror axis. The mirror symmetry is very common in large numbers of inversion asymmetric WSMs, specifically in TaAs, TaP, NbAs, and NbP family\cite{Weng-PRX15,Lee-PRB15}. At low energy the minimal model can be written as a sum of four effective Weyl Hamiltonians, $\mathcal{H}_{W}=\sum_{\gamma}\sum_{k}\Psi^\dagger_{\gamma,\mathbf{k}}H_{\gamma}(\mathbf{k})\Psi_{\gamma,\mathbf{k}}$ with $\gamma=1,2,3,4$ level the four Weyl nodes and $\Psi_{\gamma,\mathbf{k}}$ are the spinor wavefunctions. The nodes $1$ and $3$ carry positive chirality while nodes $2$ and $4$ carry negative chirality. Pair of opposite chiral nodes i.e., $\gamma=1,3$ and   $\gamma=2,4$ are related to each other by a mirror plane perpendicular to the $z$-axis. The linearized Hamiltonian at each Weyl nodes are given (taking $\hbar=1$ and $v_F=1$),
\begin{eqnarray}
\label{tilt-hamil1}
H_{1(3)}(\mathbf{k})&=&\mathcal{C}_{1(3)}k_z\sigma_0+(k_x\sigma_x+k_y\sigma_y+k_z\sigma_z)\\
H_{2(4)}(\mathbf{k})&=&\mathcal{C}_{2(4)}k_z\sigma_0+(k_x\sigma_x+k_y\sigma_y-k_z\sigma_z)
\label{tilt-hamil2}
\end{eqnarray}

We consider the tilting is along the transport direction (i.e., along the $z$-axis). We consider two types of tilts, time-reversal symmetric (TRS) and TRS broken. The TRS tilt coefficients $\mathcal{C}_i$ of positive and negative chiral Weyl nodes are related to each other by $\mathcal{C}_1=-\mathcal{C}_3$ and $\mathcal{C}_2=-\mathcal{C}_4$, respectively. On the other hand, TRS is broken if the tilt coefficients in each chirality sectors are equal i.e., $\mathcal{C}_1=\mathcal{C}_3$ and $\mathcal{C}_2=\mathcal{C}_4$. Additionally, the mirror symmetry: $R_z H(k_x,k_y,-k_z)R_z=H(k_x,k_y,k_z)$, relates the tilt coefficients of opposite chiral Weyl nodes by  $\mathcal{C}_2=-\mathcal{C}_1$ and $\mathcal{C}_4=-\mathcal{C}_3$, in both cases. $R_z$ is the reflection operator about $xy$-plane. From now, we define the magnitude of tilt in each Weyl node by $\mathcal{C}$. The Hamiltonians in each chirality sectors are invariant under TR operator $\mathcal{T}$, i.e. $[\mathcal{H}_{\pm}(\vec{r}),\mathcal{T}]=0$, in presence of TRS tilt. The TR-operator reads $\mathcal{T}=-i\sigma_y\mathcal{K}$, where $\mathcal{K}$ denotes the complex conjugation. Here, the positive and negative chiral Hamiltonians are defined as, $\mathcal{H}_{+}=diag\{H_1,H_3\}$ and $\mathcal{H}_{-}=diag\{H_2,H_4\}$. For an inversion symmetry broken WSM, there exists an emergent symmetry operation $\mathcal{U}$ which connect the two chirality sectors by \cite{Trauzettel-PRL18}, i.e., $\mathcal{U}\mathcal{H}_{\pm}(\vec{r})\mathcal{U}^{-1}=\mathcal{H}_{\mp}(\vec{r})$, in absence of tilt.  The operator $\mathcal{U}$ call the $\mathbb{Z}_2$ exchange symmetry and the operator reads $\mathcal{U}=e^{i\pi\sigma_y\tau_yR_x/2}$, where $R_x$ is the reflection operator about $yz$ plane. Now it is easy to check that, in presence of mirror symmetry, the TRS tilt preserves the $\mathbb{Z}_2$ symmetry whereas the TRS broken tilt breaks this symmetry. In appendix, we show that the tilting term appears with an energy-dependent shift $\Delta k_z$ in the momentum along transport direction, where $\Delta k_z=2\mathcal{C}\mathcal{E}/(1-\mathcal{C}^2)$. The shift has an opposite sign in two Weyl nodes of the same chirality, in case of TRS tilt. This is because the TRS tilt breaks the valley degeneracy. In presence of TRS broken tilts, the shift has opposite signs in the Weyl nodes of opposite chirality which is consequence of the $\mathbb{Z}_2$ symmetry breaking.

We consider an inversion asymmetric WSM sandwitch between two s-wave superconducting WSM. The superconductivity in WSM regions can be proximity induced by superconducting electrode. The BdG Hamiltonian for positive chirality sector is given by,
\begin{eqnarray}
\mathcal{H}^{+}_{BdG}(\phi)=\begin{pmatrix}
\mathcal{H}_{BdG}(\phi) & \emptyset\\
\emptyset & \mathcal{H}'_{BdG}(\phi)
\end{pmatrix}
\label{BdG-hamilp}
\end{eqnarray}
in the Nambu basis $(\Psi_{1,\uparrow},\Psi_{1,\downarrow},\Psi^\dagger_{3,\downarrow},-\Psi^\dagger_{3,\uparrow},\Psi_{3,\uparrow},\Psi_{3,\downarrow},\nonumber\\\Psi^\dagger_{1,\downarrow},-\Psi^\dagger_{1,\uparrow})$. Here, spin indices $\uparrow$,$\downarrow$ and $\Psi_{\gamma,\sigma}$'s are annhilation operators. The $\emptyset$ in the above equation is $4\times 4$ null matrix. The BdG Hamiltonian in Eq.(\ref{BdG-hamilp}) decouples into two blocks. The diagonal Hamiltonian part $\mathcal{H}_{BdG}(\phi)$ is given by,
\begin{eqnarray}
\mathcal{H}_{BdG}(\phi)&=&-i\mathcal{C}\partial_z\sigma_0\nu_0-\mu(\mathbf{r})\sigma_0\nu_z-i\partial_{\mathbf{r}}\cdot \mathbf{\sigma}\nu_z\nonumber\\&&+\Delta_s(\mathbf{r})e^{i\nu_zsgn(z)\phi/2}\sigma_0\nu_x
\label{diag-hamil1}
\end{eqnarray}
The $2\times 2$ Pauli matrices $\nu_i$ ($i=x,y,z,0$) acting on particle-hole space. Here, $\Delta_s(\mathbf{r})=\Delta\Theta(|z|-L/2)$ is the pairing potential, $\phi$ is the phase difference and $L$ is the junction length; $\Theta(z)$ is the Heaviside function and $sgn(z)$ is the sign function. The chemical potential is taken as: $\mu(\mathbf{r})=\mu_S\Theta(|z|-L/2)+\mu_N\Theta(L/2-z)$. The Hamiltonian $\mathcal{H}'_{BdG}(\phi)$ can be obtained by replacing $\mathcal{C}$ by $-\mathcal{C}$ in Eq.(\ref{diag-hamil1}), in case of TRS tilt. Thus the two blocks in Eq.(\ref{BdG-hamilp}) become different i.e., $\mathcal{H}_{BdG}(\phi)\neq\mathcal{H}'_{BdG}(\phi)$. However, these two BdG Hamiltonians become identical in case of TRS broken tilt. Similarly, the BdG Hamiltonian of negative chirality sector $\mathcal{H}^-_{BdG}(\phi)$ is obtained by using the Hamiltonian in Eqs.(\ref{tilt-hamil1},\ref{tilt-hamil2}). The BdG Hamiltonians of two chirality sectors are identical in case of TRS tilt because it preserves the $\mathbb{Z}_2$ exchange symmetry. They become different only in presence of TRS broken tilt(see Appendix).

\begin{figure}
\includegraphics[width=1.67in]{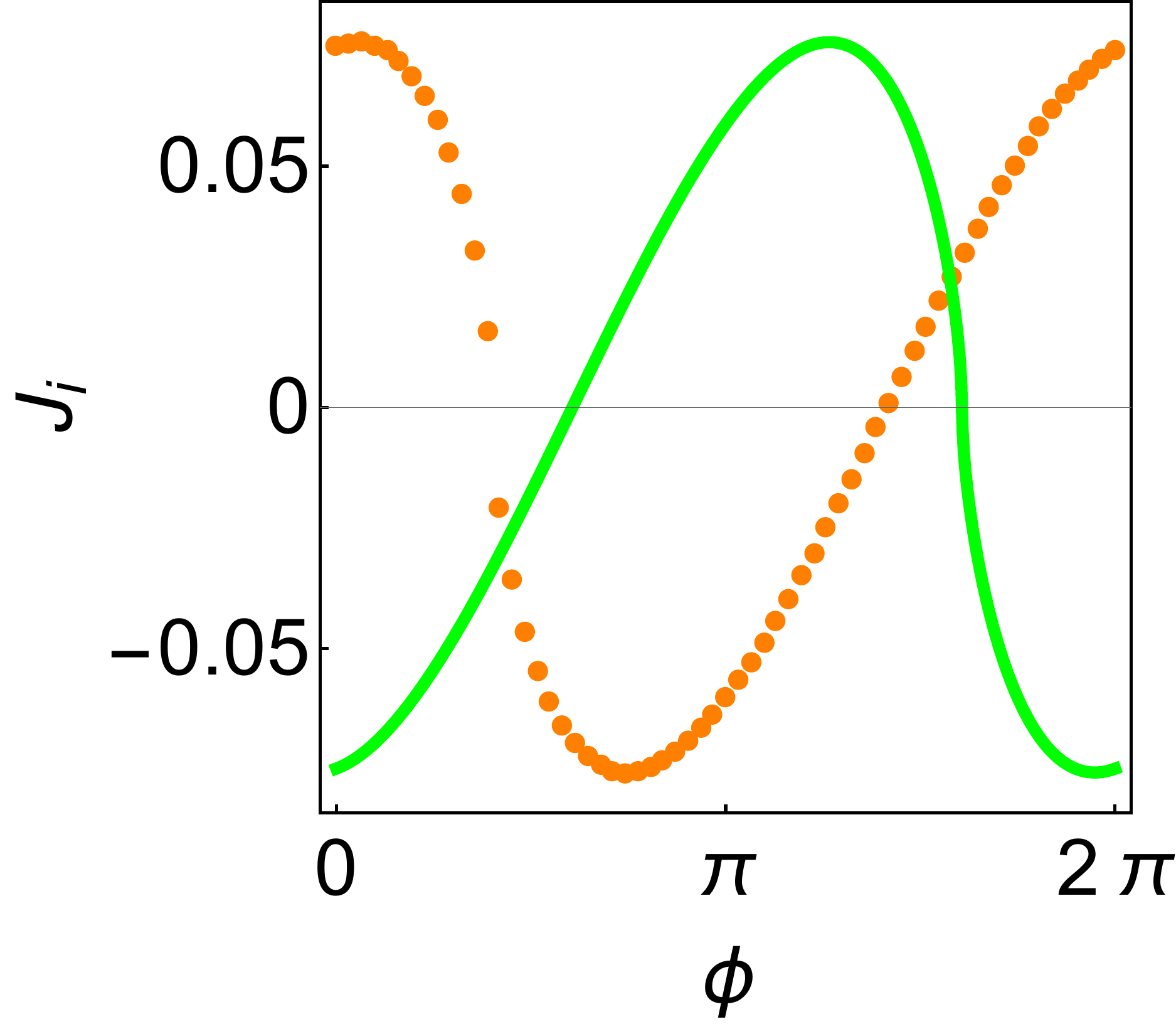}
\includegraphics[width=1.67in]{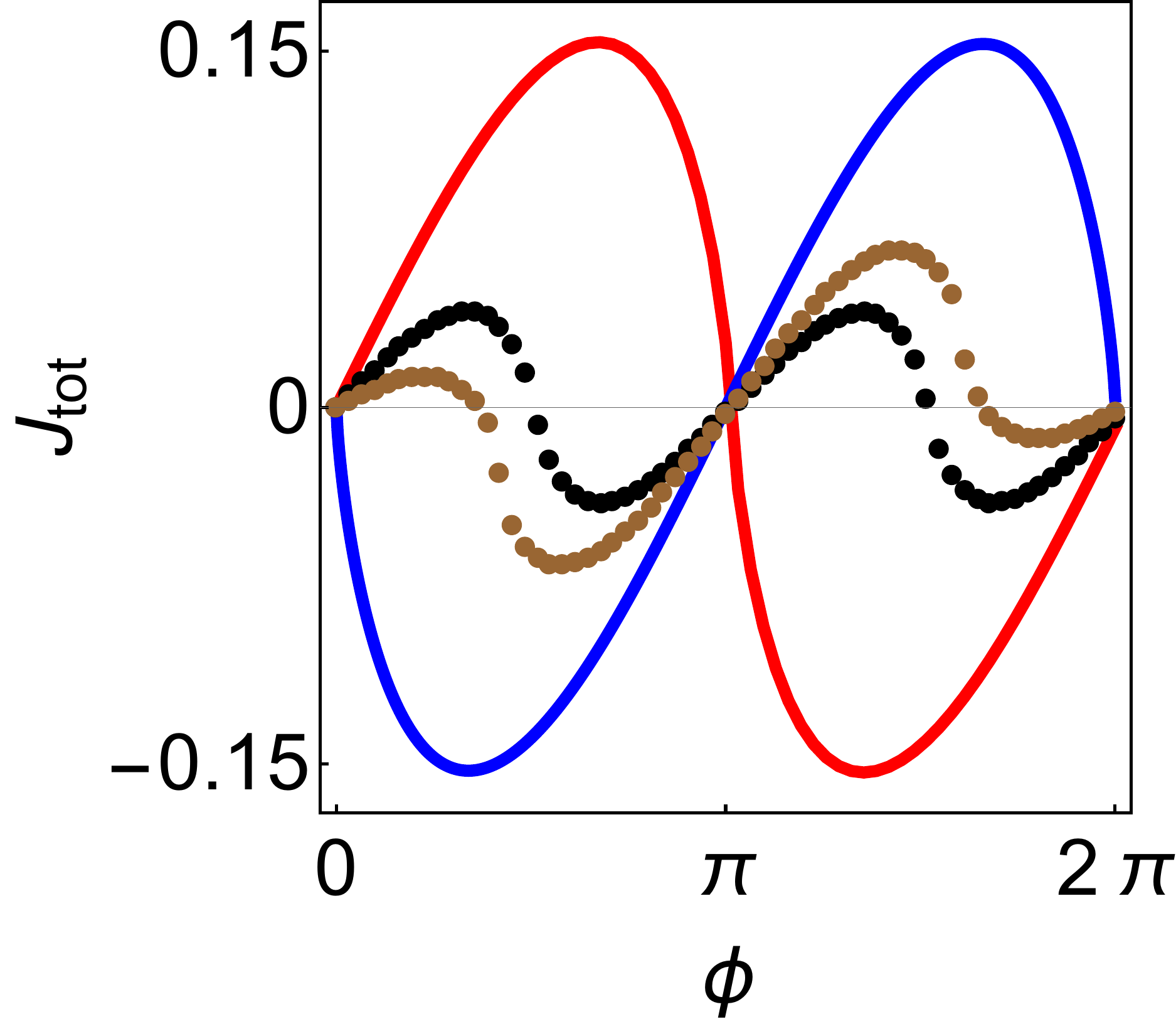}
\caption{CPR of individual and total current. Josephson current as a function of superconducting phase difference $\phi$. We fix $C=0.1$, $\mu_N/\Delta=100$ and current is expressed in unit of $e\mu^2_NW^2\Delta/\hbar$. Left panel displays the individual components of Josephson current with $L/\xi=0.03\pi$. Right Panel displays the total Josephson current. The red and blue solid lines correspond to $L/\xi=0.1\pi$ and $L/\xi=0.05\pi$. These values of $L/\xi$ satisfying $\phi_t=2n\pi$ and $\phi_t=(2n+1)\pi$, respectively. The brown purple dotted lines correspond to $L/\xi=0.07\pi$. The second harmonic Josephson current is shown by the black dotted line which corresponds to the phase $\phi_t=(2n+1)\pi/2$.}
\label{pos-chi-curr}
\end{figure}

The energy dispersions relation in superconducting region are calculated by diagonalizing the Hamiltonian in Eq.(\ref{diag-hamil1}), and given by,
\begin{eqnarray}
\mathcal{E}_s=\mathcal{C}q_z \pm \sqrt{\Delta^2_0+(\mu_s \pm \mathbf{q})^2}
\end{eqnarray}
In the normal region, the energy dispersions are obtained by diagonalizing the Hamiltonians of electron and hole separately. The dispersion relations are given by,
\begin{eqnarray}
\mathcal{E}_{e(h)}=\pm \sqrt{q^2_z+q^2_p}+\mathcal{C}q_z-(+)\mu
\label{quasi-spectra}
\end{eqnarray}
where the subscripts $e(h)$ denote the electrnlike (holelike) excitation spectra. The propagating wavevectors of quasiparticles with incident energy $E$ and transverse momentum $k_p$ are obtained from Eq.(\ref{quasi-spectra}) and are given by,
\begin{eqnarray}
q_{ep(m)}&=&\frac{-\mathcal{C}(\mu_N+E)+(-)\sqrt{(\mu_N+E)^2-(1-\mathcal{C}^2)q^2_p}}{(1-\mathcal{C}^2)}\nonumber\\
q_{hp(m)}&=&\frac{\mathcal{C}(\mu_N-E)-(+)\sqrt{(\mu_N-E)^2-(1-\mathcal{C}^2)q^2_p}}{(1-\mathcal{C}^2)}
\label{wave-vectors}
\end{eqnarray}
with $q_p=\sqrt{q^2_x+q^2_y}$. The electron and hole wavefunctions in the normal region read,
\begin{eqnarray}
\Psi^e_{in(out)}&=&e^{iq_{ep(m)}z}\begin{pmatrix} 1 & \mathcal{Q}_{ep(m)} & 0 & 0 \end{pmatrix}\\
\Psi^h_{in(out)}&=&e^{iq_{hp(m)}z}\begin{pmatrix}  0 & 0 & 1 & \mathcal{Q}_{hp(m)}\end{pmatrix}
\end{eqnarray}
where,
\begin{align}
\mathcal{Q}_{ep(m)}=\frac{q_pe^{i\theta}}{q_{\pm}+q_{ep(m)}}; \mathcal{Q}_{hp(m)}=\frac{q_pe^{i\theta}}{q'_{\pm}+q_{hp(m)}}
\end{align}
with $q_{\pm}=\sqrt{q^2_p+q^2_{ep(m)}}$ and $q'_{\pm}=\sqrt{q^2_p+q^2_{hp(m)}}$. To calculate the Josephson current in the junction is to first obtain the energy spectrum for the Andreev bound states in the normal region. This is done by matching the wavefunctions at the two SN interfaces ($z=0$ and $z=L$) in the junction and solving the allowed energy values\cite{Linder-PRB09,Kulikov-PRB20}. The boundary conditions lead to an $8\times 8$ matrix $\mathcal{M}$ for the above eight scattering coefficients. The condition $det (\mathcal{M})=0$ initiate the non-trivial relation between the Andreev bound state $\mathcal{E}$ with the superconducting phase difference $\phi$. In the short junction limit($L\ll \xi=1/\Delta$), the discrete energy spectrum of ABS comes with energy $\pm\mathcal{E}$. These two states have opposite fermion parity. In the case of TRS tilt the expression of $\mathcal{E}$ is given by:\begin{eqnarray}
\mathcal{E}=-\Delta\sqrt{1-\mathcal{C}^2}\sqrt{\frac{\mathcal{B}}{\mathcal{A}}+\frac{\mathcal{F}}{\mathcal{A}}\sin^2\frac{(\phi+\phi_t)}{2}}
\label{abs-1}
\end{eqnarray}
in which the epressions of $\mathcal{B}$, $\mathcal{A}$ and $\mathcal{F}$ are written explicitly as,
\begin{eqnarray}
\mathcal{B}&=&(\mathcal{Q}_{ep}\mathcal{Q}_{hm}+\mathcal{Q}_{em}\mathcal{Q}_{hp})\nonumber\\&&-(\mathcal{Q}_{ep}\mathcal{Q}_{hp}+\mathcal{Q}_{em}\mathcal{Q}_{hm})\cos^2(\Delta q L/2)\nonumber\\&&+\frac{(\mathcal{Q}_{hp}\mathcal{Q}_{hm}+\alpha^4\mathcal{Q}_{ep}\mathcal{Q}_{em})}{\alpha^2}\sin^2(\Delta q L/2)\nonumber\\
\mathcal{F}&=&(\mathcal{Q}_{ep}-\mathcal{Q}_{em})(\mathcal{Q}_{hp}-\mathcal{Q}_{hm})\nonumber\\
\mathcal{A}&=&(\mathcal{Q}_{em}\mathcal{Q}_{hp}+\mathcal{Q}_{ep}\mathcal{Q}_{hm})\nonumber\\&&-(\mathcal{Q}_{em}\mathcal{Q}_{hm}+\mathcal{Q}_{ep}\mathcal{Q}_{hp})\cos(\Delta q L)
\label{funt-form}
\end{eqnarray}
with $\Delta q=(q_{ep}-q_{em})$. The expression of phase $\phi_t$ is, $\phi_t=2\mu_N\mathcal{C}L/(1-\mathcal{C}^2)$. Two energy states $\pm \mathcal{E}$, crosses at phase $\phi+\phi_t=\pi$. Similarly, the discrete energy spectrum of BdG Hamiltonian $\mathcal{H}'_{BdG}$ consists the energy eigenvalues $\pm\mathcal{E}'$, which is obtained by replacing $\mathcal{Q}_{ep}\leftrightarrow \mathcal{Q}_{hp}$, $\mathcal{Q}_{em}\leftrightarrow \mathcal{Q}_{hm}$, $\alpha \rightarrow 1/\alpha$, $\phi_t\rightarrow -\phi_t$. The expresseion of $\mathcal{E}'$ is given by,
\begin{eqnarray}
\mathcal{E}'=-\Delta\sqrt{1-\mathcal{C}^2}\sqrt{\frac{\mathcal{B}}{\mathcal{A}}+\frac{\mathcal{F}}{\mathcal{A}}\sin^2\frac{(\phi-\phi_t)}{2}}
\label{abs-2}
\end{eqnarray} 
Here, the ABSs of the two chirality sectors are equal because of the $\mathbb{Z}_2$ symmetry(see Appendix). The eigenvalues $\pm\mathcal{E}$ of BdG Hamiltonian contribute the Josephson current by the following relation, \begin{eqnarray} I(\phi)=-\frac{2e}{\hbar}{\partial \mathcal{E}\over \partial \phi}f(\mathcal{E}) \end{eqnarray}
where $f(\mathcal{E})$ is the Fermi-Dirac distribution function. The supercurrents from the two branches (opposite parity) are in equal magnitude but differ in sign. Therefore, it is sufficient to consider only the Josephson current from one branch. The Josepson current of the system is obtained by integrating out the transverse momentum,
\begin{eqnarray}
J(\phi)=\frac{W^2}{(2\pi)^2}\int I(\phi) dq_x dq_y
\end{eqnarray}
where, $W$ is the dimension in both $x$ and $y$-direction. We define the total Josephson current as,
\begin{eqnarray}
J_{tot}(\phi)=J(\phi)+J'(\phi)
\label{tot-jos}
\end{eqnarray}
and valley Josephson current as,
\begin{eqnarray}
J_{valley}(\phi)=J(\phi)-J'(\phi)
\label{valley-jos}
\end{eqnarray}
where, $J(\phi)$ and $J'(\phi)$ are the Josephson currents derived respectively from Eq.(\ref{abs-1}) and Eq.(\ref{abs-2}). Contrarily, in case of TRS broken tilt, the BdG Hamiltonians $\mathcal{H}_{BdG}$ and $\mathcal{H}'_{BdG}$ of a given chirality sector, are equal which implies the energies $\mathcal{E}(\phi)$ and $\mathcal{E}'(\phi)$ are degenerate. However, in this situation the ABSs of positive ($\mathcal{E}^+(\phi)$) and negative ($\mathcal{E}^-(\phi)$) chiralty sectors become different as a consequence of $\mathbb{Z}_2$ symmetry breaking. By straightforward algebra we get, $\mathcal{E}^+(\phi)=\mathcal{E}(\phi)$ and $\mathcal{E}^-(\phi)=\mathcal{E}'(\phi)$. So, the total Josephson current here $J_{tot}(\phi)=J^{+}(\phi)+J^{-}(\phi)$, has same expression as in Eq.(\ref{tot-jos}). The chirality Josephson current is defined: $J_{chi}(\phi)=J^{+}(\phi)-J^{-}(\phi)$ and again has same expression as in Eq.(\ref{valley-jos}). However, in absence of mirror symmetry,  the currents $J_{chi}(\phi)$ and $J_{valley}(\phi)$ will be different. Importantly, the Josephson currents $J_{valley}(\phi)$ and $J_{chi}(\phi)$ vanishes in absence of TRS tilt and TRS breaking tilt, respectively.
\begin{figure}
\includegraphics[width=2.in]{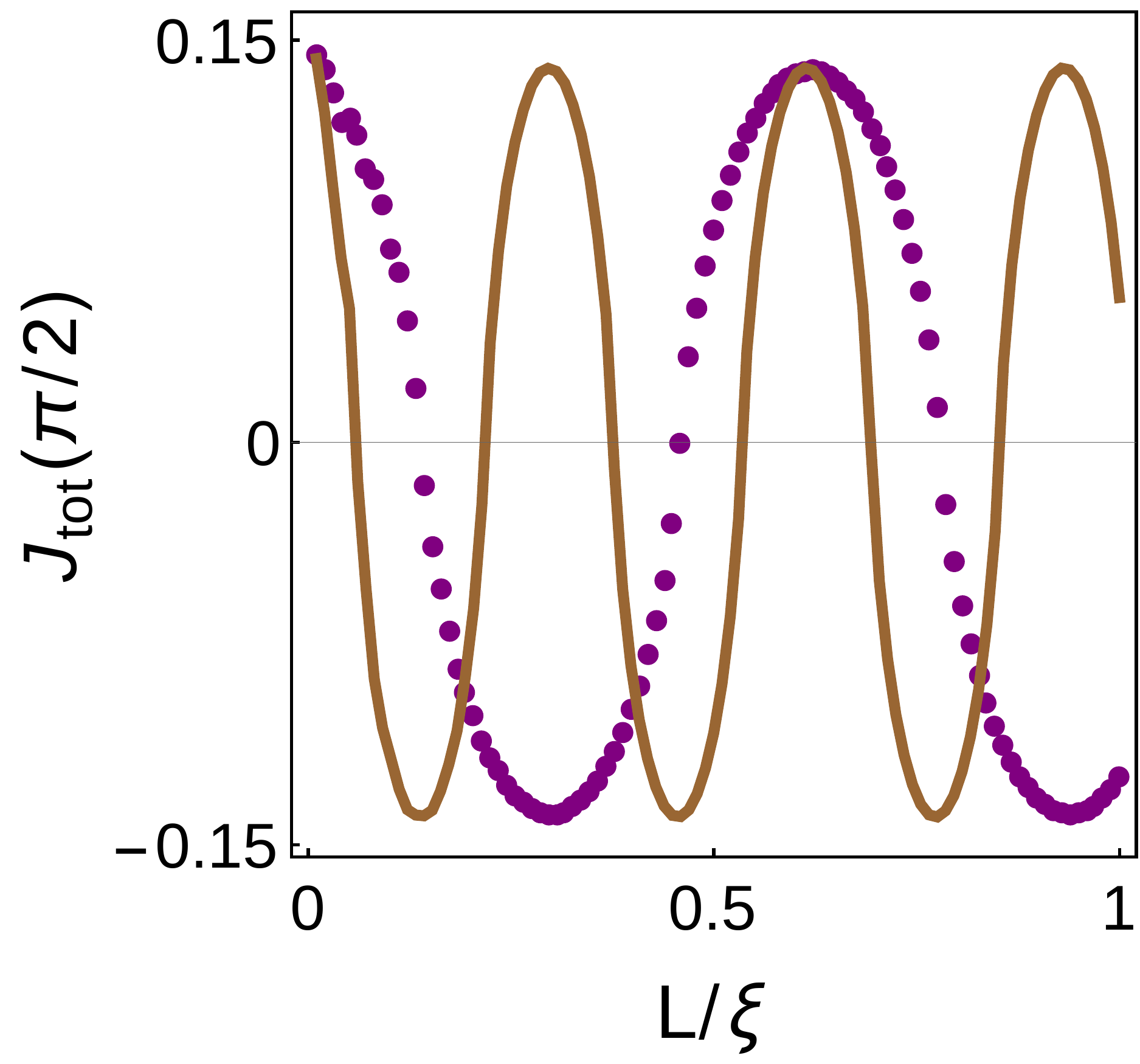}
\caption{The total Josephson current $J_{tot}$ at $\phi=\pi/2$ as a function of Josephson length with two different values of tilt. The Purple dotted line and brown solid line correspond to $\mathcal{C}=0.05$ and $.1$, respectively.}
\label{josp-length}
\end{figure}

\textbf{Total Josephson current:-}
We consider the BdG Hamiltonian of the positive chirality sector in Eq.(\ref{BdG-hamilp}) in case of TRS tilt. The TRS tilt induced phase $\phi_t$ breaks the valley degeneracy and it appears with an opposite sign in the currents $J(\phi)$ and $J'(\phi)$. Thereby these two currents become different i.e., $J(\phi) \neq J'(\phi)$. However, these two supercurrents are equal in absence of TRS tilt. The presence of phase shift $\phi_t$ leads to an anomalous Josephson current at zero- bias (i.e., $\phi=0$) $J(\phi=0)\neq 0$, $J'(\phi=0)\neq 0$. The system indeed realizes the Josephson $\phi$ junction which is controllable by the parameter $\phi_t$ i.e., by tuning the Josephson length and doping for any arbitrary value of tilt. We define an operator $\mathcal{T}_{BdG}=-i\sigma_y\mathcal{K}\tau_0$, which relate the two BdG Hamiltonians of a given chirality sector as follows: $\mathcal{T}_{BdG}\mathcal{H}_{BdG}(\phi)\mathcal{T}^{-1}_{BdG}=\mathcal{H}'_{BdG}(-\phi)$. As a result, the Andreev levels have the following symmetry: $\mathcal{E}(\phi)=\mathcal{E}'(-\phi)$ and subsequently, the current satisfies $J(\phi)=-J'(-\phi)$. Thus the total current $J_{tot} (\phi)$ is an odd function of $\phi$ ($J_{tot}(\phi)=-J_{tot}(-\phi)$). On the other hand, the TRS broken tilt induced phase $\phi_t$ breaks $\mathbb{Z}_2$ symmetry and consequently it appears with an opposite sign in the currents $J^{+}(\phi)$ and $J^{-}(\phi)$. These two currents are equal in absence of TRS broken tilt. An anomalous current also occurs at zero-bias i.e.,  $J^{+}(\phi=0)\neq 0$, $J^{-}(\phi=0)\neq 0$. The BdG Hamiltonians of two chirality sectors are related to each other by, $\mathcal{S}_{BdG}\mathcal{H}^{+}_{BdG}(\phi)\mathcal{S}^{-1}_{BdG}=\mathcal{H}^{-}_{BdG}(-\phi)$, where the operator $S_{BdG}$ is the product of $\mathcal{T}_{BdG}$ and $\mathcal{U}_{BdG}$(see Appendix). Consequently, the Andreev levels have the following symmetry: $\mathcal{E}^{+}(\phi)=\mathcal{E}^{-}(-\phi)$ and the current satisfies $J^{+}(\phi)=-J^{-}(-\phi)$. Thus, the total current $J_{tot}(\phi)=J^{+}(\phi)+J^{-}(\phi)$ is also an odd function of $\phi$. 

We have shown the current phase relation (CPR) in Fig.(\ref{pos-chi-curr}). The left panel of Fig.(\ref{pos-chi-curr}) displays the CPR of individual Josephson current components. The non-zero value of zero-bias supercurrents (when $\phi_t \neq 0$) indicates that the system is in Josephson $\phi$-junction. The zero-bias supercurrent vanishes if and only if the tilt induced phase $\phi_t=0$. Thus this zero-bias current only occurs in presence of tilt\cite{Sinha-PRB20}. The realization of anomalous Josephson currents here are different from the earlier studies\cite{Tanaka-PRB97,Buzdin-RMP05,Buzdin-PRL08,Yokoyama-PRB14,Dolcini-PRB15,tanaka-PRL09,Linder-PRL10}. The $\phi$ junction has been discussed mainly in time-reversal broken system like in presence of Zeeman field and spin-orbit coupling or in a junction of SC-Ferromagnet-SC. The right panel of Fig.(\ref{pos-chi-curr}) displays the CPR of total Josephson current ($J_{tot}(\phi)$). Since the current $J_{tot}(\phi)$ is an odd function, it always vanishes at $\phi=n\pi$. The junction is in $0$-state for $\phi_t=2n\pi$ and corresponding current $J_{tot}(\phi) \sim \sin\phi$. Increasing or decreasing the phase $\phi_t$ gradually, the amplitude of supercurrent decreases initially, and then increases eventually with the complete reversion of the Josephson current at $\phi_t=(2n+1)\pi$. The junction is in $\pi$-state for these values of $\phi_t$. Thus the Josephson current $0$ to $\pi$ transition can be realized by tuning the phase parameter $\phi_t$ both in TRS tilt and TRS broken tilt. The red and blue solid lines in Fig.(\ref{pos-chi-curr}) represent the Josephson $0$ and $\pi$-junction, respectively. The dotted curves representing the CPR other than $0$ or $\pi$ Josephson junction.
The Josephson current $0$-$\pi$ transition also evident from fig.(\ref{josp-length}). The total Josephson current at $\phi=\pi/2$ is plotted against the Josephson length for two different values of tilt. The supercurrent changes its sign with tuning the Josephson length for a fixed value of tilt and phase $\phi$. The CPR can generally be expanded in Fourier series with all the harmonics as follows: $J_{tot}(\phi)=\sum_{n}[J_n\sin n\phi+I_n\cos n\phi]$. Since the current $J_{tot}(\phi)$ in each chirality sector has time reversal symmetry, all the coefficients $I_n$ of cosine terms are zero. The leading term of $J_{tot}(\phi)$ is $\sin\phi$ and the presence of higher harmonic terms causes an extra dip or peak in the dotted curves for $\phi \in[0,\pi]$. At $\phi_t=\pi/2$, the current has $\pi$ periodicity and is dominated by the second harmonic term $\sin(2\phi)$. This is shown by black dotted line. 

\begin{figure}
\includegraphics[width=3in]{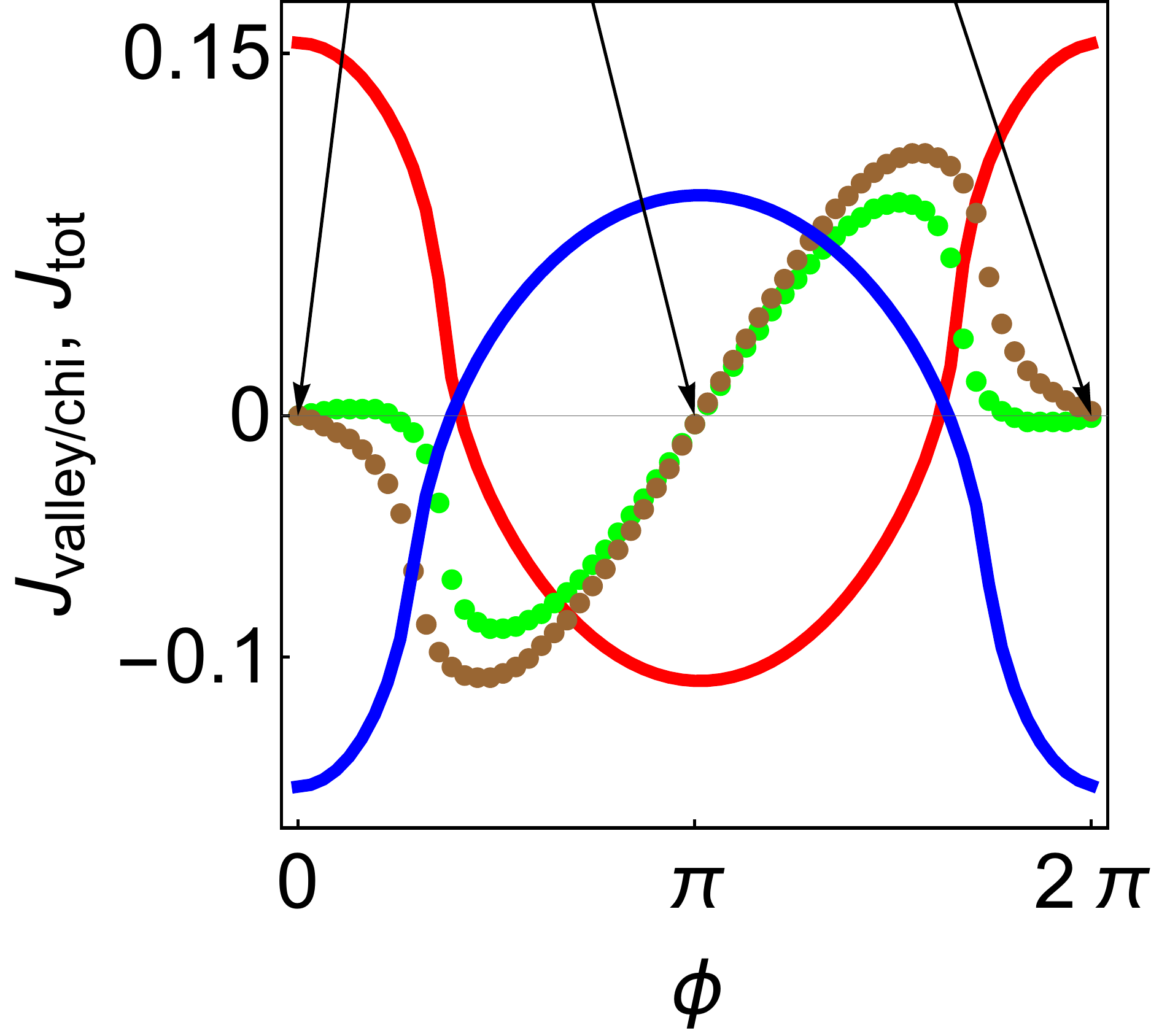}
\caption{Valley/Chirality dependent Josephson current and total Josephson current ($J_{tot}$) as a function of SC phase $\phi$. We fix $\mathcal{C}=0.1$ and $\mu_N/\Delta=100$. The red and blue solid lines display the current $J_{valley/chi}(\phi)$, whereas, the green and brown dotted lines display the current $J_{tot}(\phi)$. The red solid line and greeen dotted line corresponds to $L/\xi=.03\pi$. The blue solid line and brown dotted line corresponds to $L/\xi=.06\pi$. The arrows are indicating the points where a pure VJC/CJC occurs.}
\hspace{3pt}
\label{CPR-TRSB}
\end{figure}

\textbf{Valley/chirality Josephson current:-} 
The breaking of valley degeneracy (by TRS tilts) and $\mathbb{Z}_2$ symmetry (by TRS broken tilts) causes the striking phenomenon of finite VJC ($J_{valley}$) and finite CJC ($J_{chi}$), respectively. However, these two currents share the same expression. The valley/chirality current is an even function of $\phi$, i.e., $J_{valley/chi}(\phi)=J_{valley/chi}(-\phi)$ and has $2\pi$ periodicity in $\phi$. This signifies that the  $J_{valley/chi}(\phi)$ exists at $\phi=n\pi$ where, the $J_{tot}(\phi)$ is identically zero. Thus a pure VJC/CJC can be found at $\phi=n\pi$. This is one of the astonishing results here. It is possible to transfer a pure valley/chiral supercurrent across the junction. This could have important implications in superconductor base valleytronics/chiraliytronics. An anomalous VJC/CJC also persists at zero bias condition. Fig.(\ref{CPR-TRSB}) displays the total and valley/chirality Josephson current for two different values of $\phi_t$. The maxima of $|J_{valley/chi}(\phi)|$ and zero of $J_{tot}(\phi)$ occur respectively at $\phi=n\pi$, indicated by arrows. A pure VJC/CJC exists at these points of  $\phi$. The zero-bias VJC/CJC is an oscillatory function of $\phi_t$, given as: $J_{valley/chi}(\phi=0)\sim \sin\phi_t$. Thus, a reversible pure VJC/CJC at zero bias is feasible by appropriately choosing the phase $\phi_t$. However, these currents are reversible and are possible to occur at any value $\phi$. In Fig.(\ref{CPR-TRSB}), the red solid lines get reversed into the blue solid line by tuning the phase $\phi_t$.

\textbf{Josephson current in long junction and Quantum anomaly:-} The VJC and CJC are closely linked respectively with the valley symmetry breaking and $\mathbb{Z}_2$ symmetry breaking in this system. We now explain how the currents VJC and CJC are associated with the quantum anomaly of Cooper pairs.  In the long junction ($L\gg \xi$) and zero temperature limit, the states deep in the superconductivity gap ($\epsilon \ll \Delta$) have only significant contributions to the Josephson currents\cite{Samuelsson-PRB00}. The low energy excitations around a Fermi surface are described by the helical model with Hamiltonian,
\begin{eqnarray}
H=v_F (q_z+\frac{\mathcal{C}\mu_N}{1-\mathcal{C}^2}s_z) s_z
\label{linear-hamil}
\end{eqnarray}
with $v_{F}=(\mu^2_N-(1-\mathcal{C}^2)k^2_p)^{1/2}/\mu_N$. The Pauli matrix $s_z$ acting on two valleys in case of TRS tilt whereas it acting on two chirality sectors in case of TRS breaking tilt. The ABS spectrum of linear Hamiltonian of Eq.(\ref{linear-hamil}) is given by\cite{Beenakker-PRL13,Crepin-PRL14,Crepin-Physica16},
\begin{eqnarray}
\epsilon^\eta_{\pm}=\pm\frac{\pi v_F}{L}(n+\frac{1}{2}+\frac{\phi+\eta \phi_t}{2\pi})
\end{eqnarray}
where $\eta=\pm 1$ representing two valleys or two chirality sectors. Compare to the short junction limit, the energy levels here are unbounded and linear in $\phi \pm \phi_t$. Each excitation spectra are associated with a fermion parity number $n$. The spectrum is invariant under the transformation $\phi+\eta \phi_t \rightarrow 2\pi$ and $n+1 \rightarrow n$. Therefore, if the phase $\phi+ \eta\phi_t$ is advanced by $2\pi$, the system changes the parity as if, an extra quasiparticle has been added to (or, remove from) the system. At zero temperature, the two energy levels of opposite parity close to the Fermi energy plays the role. Once an eigenstate changes the parity, it can't relax back without changing the phase $\phi+\eta \phi_t$. Thus, the parity of an eigenstate is no longer a conserved quantity in quantum dynamics. This is referred to as fermion number parity anomaly\cite{Beenakker-PRL13,Crepin-PRL14}. The anomaly also presents in a short junction. The Josephson current is obtained by the method of Bosonization\cite{Trauzettel-PRL18,Beenakker-PRL13,Crepin-PRL14,Crepin-Physica16}. For a given transverse momentum $k_p$ the current expression is given by,
\begin{eqnarray}
j_{\eta}=\frac{2\pi v_F}{L}(\frac{\phi +\eta \phi_t}{\pi}-sgn(\phi +\eta \phi_t-\pi)-1)
\end{eqnarray}
Different from the short junction, the supercurrent has sign ambiguity and it occurs due to the change in slope of ABSs at phase $\phi+\eta \phi_t=\pi$ (i.e., where an eigenstate changes parity). Integrating out the transverse moemntum, we find the total Josephson current at $\phi=\pi$ is,
\begin{eqnarray}
J_{\eta}=\eta\frac{2\mu^2_N}{3L}(\frac{\mathcal{C}L}{\pi}-sgn(\mathcal{C}))
\label{pos-chiral-long}
\end{eqnarray}
Here, we neglect the contribution  from the term $\mathcal{C}^2$. Thus the valley/chirality Josephson current for small value of $\phi_t$ is given by,
\begin{eqnarray}
J_{valley/chi}=\frac{4\mu^2_N}{3L}[\frac{\mathcal{C}L}{\pi}-sgn(\mathcal{C})]
\label{valley/chi-long}
\end{eqnarray}
The current in Eq.(\ref{valley/chi-long}) shows a discontinuous jump when $\mathcal{C}\rightarrow 0$. It is the failure of restoring fermion parity when the symmetry breaking parameter $\phi_t$ sent to zero. The discontinuity in the current signifies that the valley and chiral Josephson current does not vanish, which contradicts the valley and $\mathbb{Z}_2$ symmetry, respectively. However, the current $J_{valley/chi}$ vanishes exactly when $\mathcal{C}=0$. The anomaly of Cooper pair is thus associated with valley/$\mathbb{Z}_2$ symmetry and somewhat analogous to the mirror anomaly in Dirac semimetal\cite{Burkov-PRL18} or parity anomaly in $2D$ Dirac fermions\cite{Semenoff-PRL84}. The quantum anomaly with $\mathbb{Z}_2$ symmetry also proposed in a Josephson junction of an IR-breaking WSM but in presence of Zeeman field\cite{Trauzettel-PRL18}. However, a quantum anomaly in both valley current and as well as in chirality Josephson current can be realized in presence of tilt.

\textbf{Discussions and Conclusions:-} In our theory, VJC or CJC results from the breaking of the valley or $\mathbb{Z}_2$ symmetry of Weyl Hamiltonian by the tilt of the Weyl cones. Therefore, our results are equally valid for spinless WSMs\cite{Okugawa-PRB17,Kim-PRB16}. Our theory could be verified in real WSMs with inversion symmetry breaking, for examples TaAs, TaP, NbAs, and NbP\cite{Weng-PRX15,Lee-PRB15}. Among these, TaAs class of materials are identified as type-I Weyl semimetals and have linear energy dispersion in wide frequency ranges. 

We now discuss the relevancy of our theory to experiments. Considering, Fermi velocity of electrons $v_F=10^6$ m/s, $\mathcal{C}/v_F=0.1$ and $\mu_N/\Delta=100$, the position shift of Weyl nodes is $\Delta k_{valley/chi}=2\times 10^{-3}$nm$^{-1}$. We find the tilt induced phase shift $\phi_t=0.8$, for a junction length $L=400$ nm. Thus, the phase has substantial value even for a small tilt.

In summary, we find the tilting can lead to a Josephson current $0$-$\pi$ transition and Josephson $\phi$ junction in an inversion-asymmetric WSM Josephson junction. Importantly, these anomalous effects can be realized in time-reversal invariant WSMs, also. The TRS tilts give rise to a pure VJC whereas, TRS breaking tilts produce a finite pure CJC which can exist even at zero bias condition. Consequently, in the long junction and at zero temperature, VJC and CJC are associated with a quantum anomaly of Cooper pairs. A pure VJC and CJC reversible can be realized by tuning the phase $\phi_t$, which means the reversible can occur by tuning the length $L$ or doping $\mu_N$ for a fixed value of tilt $C$ of Weyl nodes. The mechanism of phase shift in the Josephson current here is completely different from the conventional ferromagnetic Josephson junctions. The spin polarization is essential in the latter case.

\textbf{Acknowledgments:-} I would like to thank S. A. Jafari and C. Beenakker for useful discussions.

\appendix

\section{Analogy with Ferromagnetic Josephson Junction}
The Hamiltonian of a single Weyl node in presence of tilt is given by,
\begin{eqnarray}
H(\mathbf{k})=\mathcal{C}k_z\sigma_0+(k_x\sigma_x+k_y\sigma_y+k_z\sigma_z)
\label{tilt-Hamil-pos}
\end{eqnarray}
The equienergy surface equations of the above Hamiltonian are given by,
\begin{eqnarray}
[k_z-\frac{\mathcal{C}\mathcal{E}}{(1-\mathcal{C}^2)}]^2+\frac{k^2_x}{(1-\mathcal{C}^2)}+\frac{k^2_y}{(1-\mathcal{C}^2)}=\frac{\mathcal{E}^2}{(1-\mathcal{C}^2)^2}\nonumber\\
\end{eqnarray}
So, the Fermi surfaces are ellipsoids cetered at $(\mp \frac{\mathcal{C}_+\mathcal{E}}{(1-\mathcal{C}^2_+)},0,0)$. The following transformation,
\begin{eqnarray}
k'_z&=&k_z-\frac{\mathcal{C}\mathcal{E}}{(1-\mathcal{C}^2)}, k'_x=\frac{k_x}{\sqrt{1-\mathcal{C}^2}}, k'_y=\frac{k_y}{\sqrt{1-\mathcal{C}^2}},\nonumber\\ &&\mathcal{E}'=\frac{\mathcal{E}}{(1-\mathcal{C}^2)}
\label{cord-trans}
\end{eqnarray}
transform the surface to a sphere in the new coordiations,
\begin{eqnarray}
\mathcal{E'}^2=k'^2_z+k'^2_x+k'^2_y
\end{eqnarray}

\begin{figure}
\includegraphics[width=1.67in]{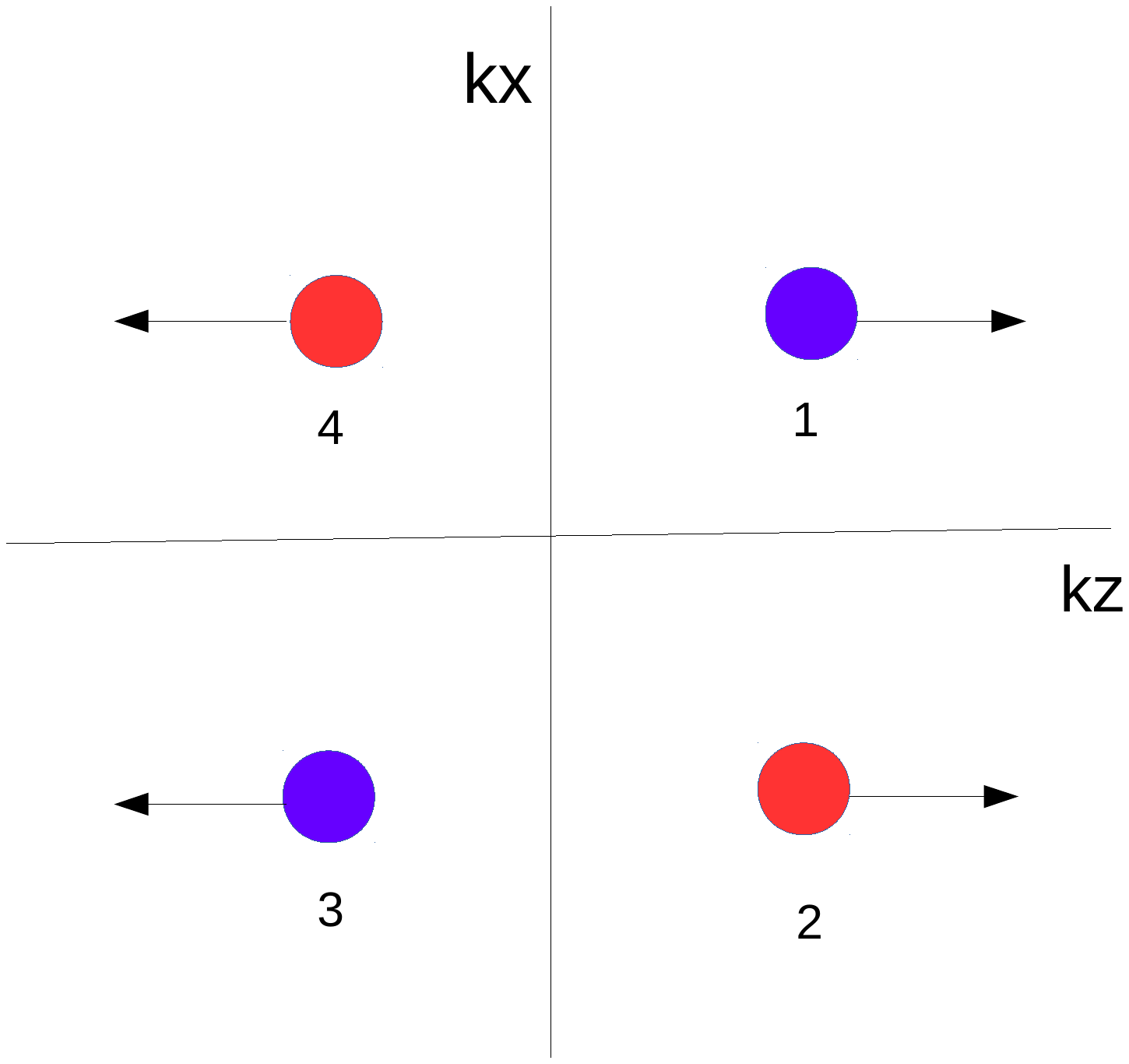}
\includegraphics[width=1.67in]{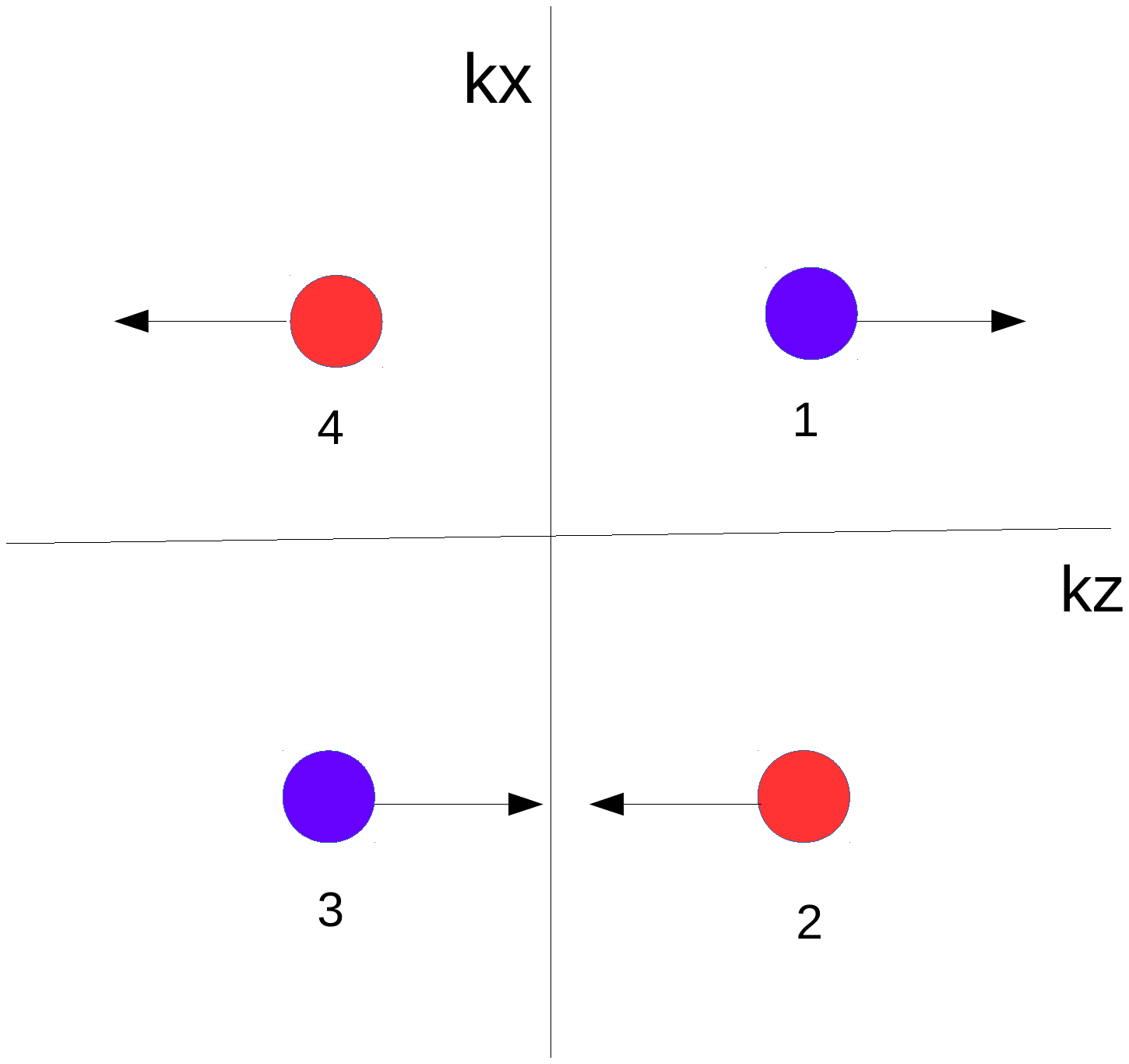}
\caption{Momentum shift due to tilt of a inversion asymmetric Weyl nodes with mirror symmetry. Blue and red circles indicates Weyl nodes of positive and negative chiralities, respectively. Right (left) panel shows the moementum shift of Weyl nodes in case of TRS (TRS broken) tilt.}
\hspace{3pt}
\label{sch-fig}
\end{figure}

From Eq.(\ref{cord-trans}), it is clear that the tilting term renormalized a quasiparticle momenta with an energy-dependent shift in the momentum along the transport direction. For a TRS tilt, the momentum of time-reversal pair Weyl nodes are shifted oppositely. Thus, a pair of electrons at the Fermi surface acquire a net valley momentum, $\Delta k_{valley}=2\mathcal{C}\mathcal{E}/(1-\mathcal{C}^2)$, at normal incidence. Similarly, the shifting of momentum occurs in the negative chirality sector with an equal magnitude. In contrast, the momentum of opposite chiral electrons is shifted in opposite directions in the case of TRS broken tilt. However, the shifting is equal and opposite in presence of a mirror symmetry which results in a net chiral momentum, $\Delta k_{chi}=2\mathcal{C}\mathcal{E}/(1-\mathcal{C}^2)$. This is akin to the ferromagnetic Josephson junction where the center of mass momentum originated via the spin splitting\cite{Trauzettel-PRL18}. We have shown the momentum shift of Weyl nodes both TRS and TRS broken tilt in Fig.(\ref{sch-fig}). Blue and red circles are representing positive and negative chiral Weyl nodes. The positive chiral nodes $1$ and $3$ are shifted oppositely by $\Delta k_{valley}$, in momentum space in presence of TRS tilt. Similarly, the negative chiral nodes $2$ and $4$ are shifted oppositely by $\Delta k_{valley}$. On the other hand, the positive and negative chiral nodes are shifted oppositely by $\Delta k_{chi}$ in momentum space in case of TRS broken tilt. The net chiral momentum and valley momentum shift give rise to CJC and VJC in a Josephson junction involving inversion asymmetric WSMs. 

\section{Scattering Wavefunctions, Andreev Bound states, and Josephson current}

\subsection{TRS tilt}
We consider the BdG Hamiltonian of positive chirality sector in presence of TRS tilt. The matrix form of the Hamiltonian is given,
\begin{align}
\mathcal{H}^{+}_{BdG}(\phi)=\begin{pmatrix}
\mathcal{H}_{BdG}(\phi) & \emptyset\\
\emptyset & \mathcal{H}'_{BdG}(\phi)
\end{pmatrix}
\label{BdG-hamilp-supp}
\end{align}
The matrix form of $\mathcal{H}_{BdG}$ is written as,
\begin{align}
\mathcal{H}_{BdG}(\phi)=
\begin{pmatrix}
H_1(\mathbf{k})-\mu & \Delta(\phi)\\
\Delta^{*}(\phi) & \mu-\mathcal{T} H_{1}(\mathbf{k}) \mathcal{T}^{-1}
\end{pmatrix}
\label{bdg-matrix-supp}
\end{align}
where $\Delta(\mathbf{r})=|\Delta|e^{i sgn(z)\phi/2}$. Here, we consider that doping $\mu_s$ in superconducting region to be large i.e., $\mu_S \gg \mu_N$. The diagonalization of Eq.(\ref{bdg-matrix-supp}) yields the eigenvalues,
\begin{eqnarray}
\mathcal{E}_{s}=\mathcal{C}k_z \pm\sqrt{\Delta^2+(\mu_s \pm k_z)^2}
\label{energy-spec-super-1}
\end{eqnarray}
where first $+(-)$ sign denotes the electronlike (holelike) excitations while second $+(-)$ sign denotes the conduction (valance) band. In the two superconducting regions, the basis functions are,
\begin{eqnarray}
\phi_{qe+}(z)&=&{\begin{pmatrix} \alpha e^{i\beta_1} & 0 & e^{-i\phi_t} & 0 \end{pmatrix}}^Te^{ik^+_{se}z}\nonumber\\
\phi_{qe-}(z)&=&{\begin{pmatrix} 0 & \frac{e^{i\beta_2}}{\alpha}& 0 &  e^{-i\phi_t} \end{pmatrix}}^T e^{ik^-_{se}z}\nonumber\\
\phi_{qh+}(z)&=&{\begin{pmatrix} 0 & e^{i\phi_t} & 0 & \alpha e^{i\beta_2} \end{pmatrix}}^T e^{ik^-_{sh}z}\nonumber\\
\phi_{qh-}(z)&=&{\begin{pmatrix} e^{i\phi_t} & 0 & \frac{e^{i\beta_1}}{\alpha} & 0 \end{pmatrix}}^T e^{ik^+_{sh}z}
\end{eqnarray}
where $t \in \{R,L\}$ labels the superconducting pairing phase on the left and right hand sides $\phi_L=-\phi/2$ and $\phi_R=\phi/2$, respectively. Here,
\begin{eqnarray}
\beta_{1(2)}=\arccos\big[\frac{\mathcal{E}\mp\mathcal{C}\mu_s}{\Delta\sqrt{(1-\mathcal{C}^2)}}\big]
\end{eqnarray} 
and
\begin{eqnarray}
\alpha=\sqrt{\frac{1-\mathcal{C}}{1+\mathcal{C}}}
\end{eqnarray}

\begin{figure}
\includegraphics[width=1.6in]{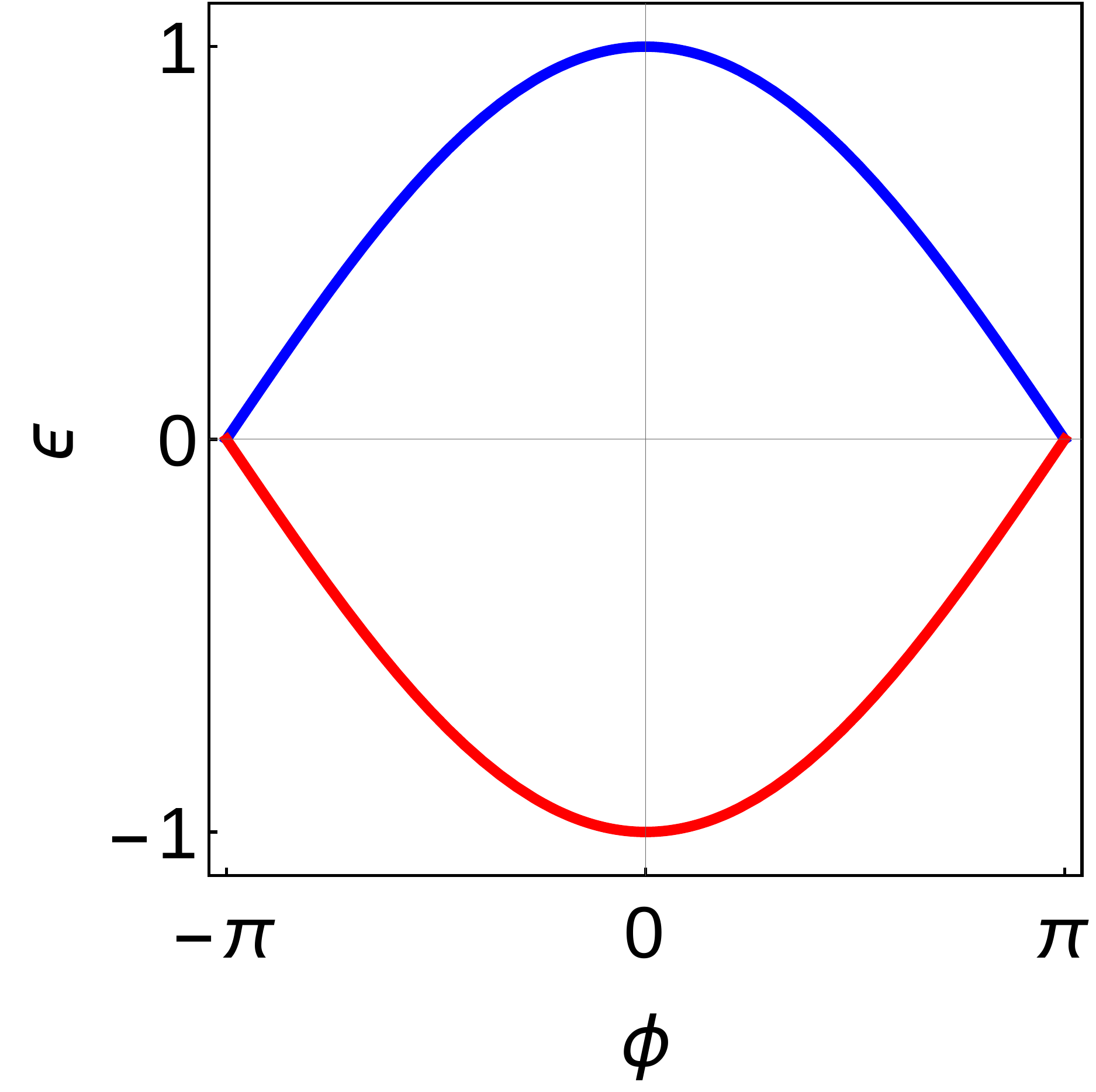}
\includegraphics[width=1.6in]{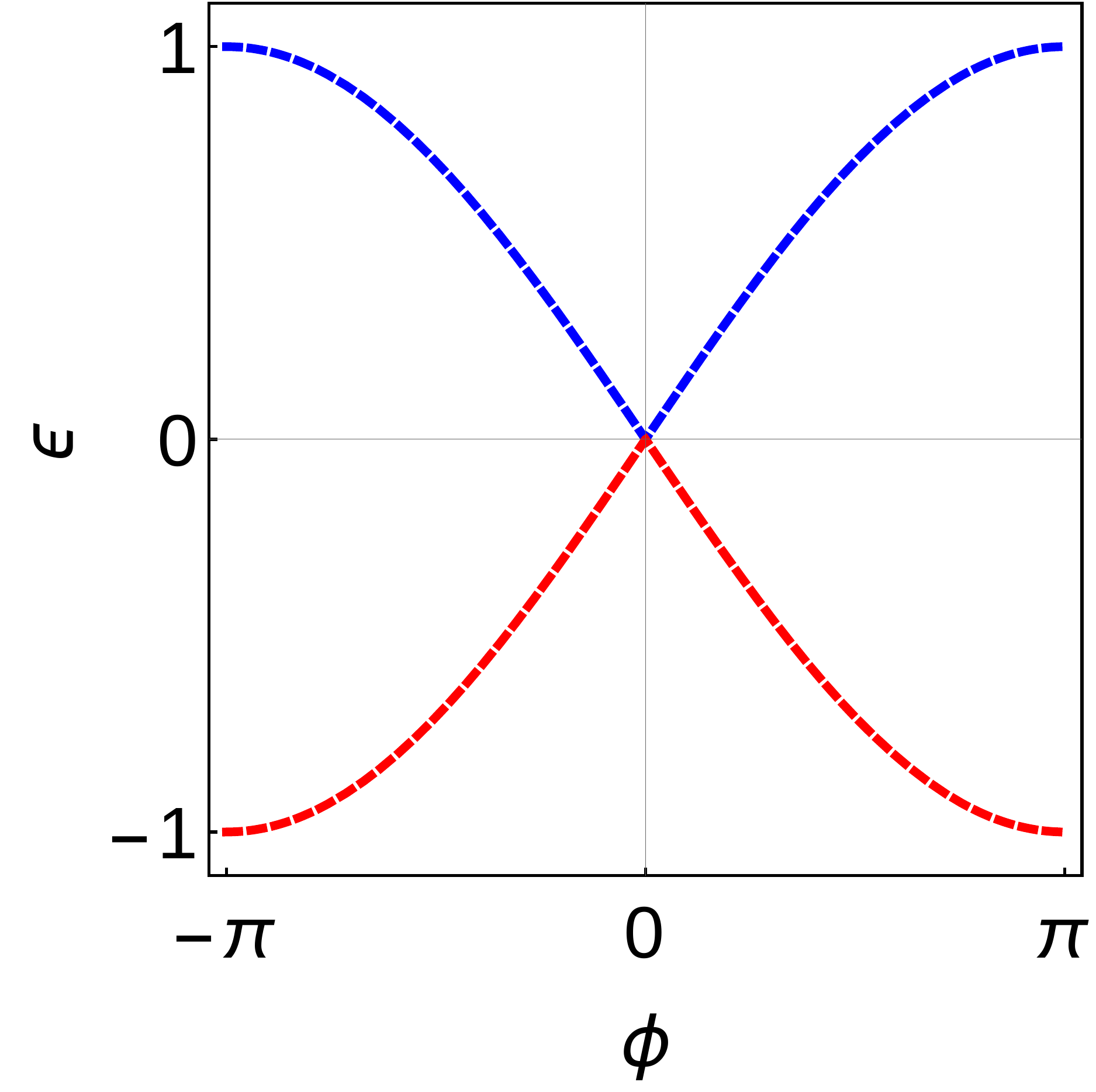}
\caption{Plot of ABSs in Eq.(\ref{abs-1-supp}) as a function of superconducting phase $\phi$. We fix $\phi_t=0$ and $\phi_t=\pi$ in right and left panel, respectively.}
\label{abs}
\end{figure}

The wavevectors $k^{\pm}_{se(h)}$ are obtained from Eq.(\ref{energy-spec-super-1}) and given below,
\begin{eqnarray}
k^{\pm}_{se}&=&\frac{\pm \mu_s-\mathcal{C}\mathcal{E}\pm \sqrt{(\mathcal{E}\mp \mathcal{C}\mu_s)^2-(1-\mathcal{C}^2)\Delta^2}}{(1-\mathcal{C}^2)}\nonumber\\
k^{\pm}_{sh}&=&\frac{\pm \mu_s-\mathcal{C}\mathcal{E}\mp \sqrt{(\mathcal{E}\mp \mathcal{C}\mu_s)^2-(1-\mathcal{C}^2)\Delta^2}}{(1-\mathcal{C}^2)}
\label{wave-vec-sup}
\end{eqnarray}
The above expressions in Eq.(\ref{wave-vec-sup}) of quasiparticles wave-vectors takes the form $k^{\pm}_{se}=\pm(\mu_s+\Omega)$ and $k^{\pm}_{sh}=\pm(\mu_s-\Omega)$, respectively for $\mathcal{C}=0$\cite{Trauzettel-PRL18}. Here, $\Omega=i\sqrt{\Delta^2-\mathcal{E}^2}$ for subgap energies $\mathcal{E}\leq \Delta$, while  $\Omega=sgn(\mathcal{E})\sqrt{\mathcal{E}^2-\Delta^2}$ for $\mathcal{E}>\Delta$. To obtain Andreev bound states, We impose the boundary conditions along $\hat{z}$, i.e., $\Psi^L_S(z=0)=\Psi_N(z=0),\Psi_N(z=L)=\Psi^R_S(z=L)$, where $\Psi^L_S(z)$, $\Psi^R_S (z)$ and  $\Psi_N(z)$ are respectively the wavefunctions in the left superconductor, right suerconductor and normal reion. The wavefunctions in the three different regions are written explicitly,
\begin{eqnarray}
\Psi^L_{S}&=&t_1\phi_{qe-}+t_2\phi_{qh-}\nonumber\\
\Psi_{N}&=&a_1\Psi^e_{in}+a_2\Psi^e_{out}+a_3\Psi^h_{in}+a_4\Psi^h_{out}\nonumber\\
\Psi^R_{S}&=&t_3\phi_{qe+}+t_4\phi_{qh+}
\end{eqnarray}
Here, $t_i$ and $a_i$ $(i=1,2,3,4)$ are the scattering amplitudes of quaiparticles (electron and hole) in three different regions. The form of different components $\Psi^e_{in(out)}$ and $\Psi^h_{in(out)}$ are given,
\begin{eqnarray}
\Psi^e_{in(out)}&=&e^{ik_{ep(m)}z}\begin{pmatrix} 1 & \mathcal{Q}_{ep(m)} & 0 & 0 \end{pmatrix}\nonumber\\
\Psi^h_{in(out)}&=&e^{ik_{hp(m)}z}\begin{pmatrix}  0 & 0 & 1 & \mathcal{Q}_{hp(m)}\end{pmatrix}
\end{eqnarray}
where,
\begin{eqnarray}
\mathcal{Q}_{ep(m)}=\frac{k_pe^{i\theta}}{k_{\pm}+k_{ep(m)}} ;\mathcal{Q}_{hp(m)}=\frac{k_pe^{i\theta}}{k'_{\pm}+k_{hp(m)}}
\end{eqnarray}
The wavevectors $k_i$ are given,
\begin{eqnarray}
k_{ep(m)}&=&\frac{-\mathcal{C}(\mu_N+E)\pm\sqrt{(\mu_N+E)^2-(1-\mathcal{C}^2)k^2_p}}{(1-\mathcal{C}^2)}\nonumber\\
k_{hp(m)}&=&\frac{\mathcal{C}(\mu_N-E)\mp\sqrt{(\mu_N-E)^2-(1-\mathcal{C}^2)k^2_p}}{(1-\mathcal{C}^2)}
\end{eqnarray}
where $k_p=\sqrt{k^2_x+k^2_y}$, $k_{+(-)}=\sqrt{k^2_p+k^2_{ep(m)}}$, $k'_{+(-)}=\sqrt{k^2_p+k^2_{hp(m)}}$ and $\theta=\tan^{-1}(k_y/k_x)$. The two boundary conditions leads to eight linear equations in the matrix form $\mathcal{M}\mathcal{X}=0$ where $\mathcal{M}$ is $8 \times 8$ matrix and $\mathcal{X}$ is the coulmn vector containing the eight scattering coefficients. The Andreev bound states are obtained by demanding the nozero solutions of these equations or equivalently from the condition $det(\mathcal{M})=0$\cite{Sinha-PRB20,Linder-PRB09,Kulikov-PRB20}. Here we take $\mu_N, \mu_S\gg\Delta$ and concentrate on the short junction limit $L \ll \xi=\hbar v/\Delta$. This allows us to neglect the contributions from the Andreev states $\mathcal{E}_b>\Delta$. In the short junction limit, the quasiparticles wave vectors are related to each other by $q_{3(4)}=-q_{1(2)}$. The condition $det(\mathcal{M})=0$ lead to the following equation,
\begin{eqnarray}
\mathcal{A}\cos^2\beta=\mathcal{B}+\mathcal{F}\sin^2\frac{(\phi+\phi_t)}{2}
\label{supp-1}
\end{eqnarray}
To obtain the above form, we take the assumption $\beta_1\simeq \beta_2=\beta=\arccos(\frac{\mathcal{E}}{\Delta\sqrt{(1-\mathcal{C}^2)}})$. We obtain the ABSs with energies $\pm \mathcal{E}$ from Eq.(\ref{supp-1}) and the analytical form is given below.
\begin{eqnarray}
\mathcal{E}=\Delta\sqrt{1-\mathcal{C}^2}\sqrt{\frac{\mathcal{B}}{\mathcal{A}}+\frac{\mathcal{F}}{\mathcal{A}}\sin^2\frac{\phi+\phi_t}{2}}
\label{abs-1-supp}
\end{eqnarray}
The analytical expressions of $\mathcal{A}$, $\mathcal{B}$ and $\mathcal{F}$ are given in Eq.(\ref{funt-form}). The two eigenvalues $\pm\mathcal{E}$ have opposite fermion number parity. We have shown the phase dependency of ABSs in Fig.(\ref{abs}). At zero temperature and $\phi_t=0$, the state $\mathcal{E}<0$ is filled up and the ground state parity is even (odd). Now, if $\phi_t$ changes to $\pi$, the state switches parity (at $\phi=0$) to odd (even).
The Josephson current at low temperature ($T\ll \Delta/k_B$, $k_B$ is the Boltzman constant) is determined from the ABSs by,
\begin{eqnarray}
I(\phi)=-\frac{2e}{\hbar}{\partial \mathcal{E} \over \partial \phi}f(\mathcal{E})
\end{eqnarray}
where $f(\mathcal{E})$ is the Fermi-Dirac distribution function. However, the current from two branches in Fig.(\ref{abs}) differ only in sign. The Josepson current is now obtain as,
\begin{eqnarray}
J(\phi)=\frac{W^2}{(2\pi)^2}\int I(\phi) dq_x dq_y
\end{eqnarray}
where, $W$ is the dimension in both $x$ and $y$-direction. We now consider the BdG Hamiltonian $\mathcal{H}'_{BdG}$ in Eq.(\ref{BdG-hamilp-supp}). The matrix form of $\mathcal{H}'_{BdG}$ is given by,
\begin{align}
\mathcal{H}'_{BdG}(\phi)=
\begin{pmatrix}
H_3(\mathbf{k})-\mu & \Delta(\phi)\\
\Delta^{*}(\phi) & \mu-\mathcal{T} H_{3}(\mathbf{k}) \mathcal{T}^{-1}
\end{pmatrix}
\label{bdg-matrix-supp1}
\end{align}
The Hamiltonian $\mathcal{H}'_{BdG}$ is thus obtained by replacing $\mathcal{C}$ by $-\mathcal{C}$ in Eq.(\ref{bdg-matrix-supp}). The eigenvalues are given by,
\begin{eqnarray}
\mathcal{E}'_{s}=-\mathcal{C}k_z \pm\sqrt{\Delta^2+(\mu_s \pm k_z)^2}
\label{energy-spec-super}
\end{eqnarray}
We similarly construct wavefunctions in three different regions by replacing $\mathcal{C}$ by $-\mathcal{C}$. This replacement effectively leads to the followings subsitution: $\mathcal{Q}_{ep}\leftrightarrow \mathcal{Q}_{hp}$, $\mathcal{Q}_{em}\leftrightarrow \mathcal{Q}_{hm}$, $\alpha \rightarrow 1/\alpha$, $\phi_t\rightarrow -\phi_t$. The analytical expressions of $\mathcal{A}$, $\mathcal{B}$ and $\mathcal{F}$ are remained unaltered. The ABSs are given,
\begin{eqnarray}
\mathcal{E}'=\Delta\sqrt{1-\mathcal{C}^2}\sqrt{\frac{\mathcal{B}}{\mathcal{A}}+\frac{\mathcal{F}}{\mathcal{A}}\sin^2\frac{(\phi-\phi_t)}{2}}
\label{abs-2-supp}
\end{eqnarray}
The ABSs of negative and positive chirality sectors are equal due to the $\mathbb{Z}_2$ exchange symmetry (see Eq.(\ref{BdG-symTRS})). The total and valley Josephson currents are now given by,
\begin{align}
J_{tot}(\phi)&=&J_1 \sin(\phi+\phi_t)+J_2 \sin(\phi-\phi_t)\nonumber\\
J_{valley}(\phi)&=&J_1 \sin(\phi+\phi_t)-J_2 \sin(\phi-\phi_t)
\end{align} 
The valley current $J_{valley}(\phi)$ vanishes when $\mathcal{C}=0$.

\subsection{TRS breaking tilt}

We now consider the BdG Hamiltonian of positive chirlity sector in presence of TRS breaking tilt. In this case the diagonal Hamiltonian in Eq.(\ref{BdG-hamilp-supp}) are same i.e., $\mathcal{H}_{BdG}(\phi)=\mathcal{H}'_{BdG}(\phi)$. However, due to $\mathbb{Z}_2$ breaking the ABSs $\mathcal{E}^+$ and $\mathcal{E}^-$ of negative and positive chirality sectors are different. The ABSs for positive and negatie chirality sectors are found to be equal $\mathcal{E}$ in Eq.(\ref{abs-1-supp}) and $\mathcal{E}'$ in Eq.(\ref{abs-2-supp}), respectively. In absence of mirror symmetry, the magntitude of tilt induced phase will be different in two chirality sectors. In this situation, it can be shown that $\mathcal{E}^+\neq \mathcal{E}$ and $\mathcal{E}^-\neq \mathcal{E}'$. The Josephson currents in this case are given by,
\begin{align}
J_{tot}(\phi)&=&J_3 \sin(\phi+\phi_t)+J_4 \sin(\phi-\phi'_t)\nonumber\\ J_{chi}(\phi)&=&J_3 \sin(\phi+\phi_t)-J_4 \sin(\phi-\phi'_t)
\end{align} 
where $\phi_t$ and $\phi'_t$ are tilt induced phase in positive and negative chirality sectors, respectively. However, we have not discuss this situation here.

\section{Symmetry Analysis of Josephson current}
At low energy and considering the mirror symmetry, the Hamiltonians of positive and negative chirality in presence of a TRS tilt, are given by,
\begin{eqnarray}
H_{+}(\mathbf{r})&=&-i\mathcal{C}\partial_z\sigma_0\tau_z-i\tau_0(\partial_x\sigma_x+\partial_y\sigma_y+\partial_z\sigma_z)\nonumber\\
H_{-}(\mathbf{r})&=&-i\mathcal{C}\partial_z\sigma_0\tau_z-i\tau_0(\partial_x\sigma_x+\partial_y\sigma_y-\partial_z\sigma_z)\
\label{TRS-hamil}
\end{eqnarray}
whereas for a TRS broken tilt the above Hamiltonians are written as
\begin{eqnarray}
H_{+}(\mathbf{r})&=&-i\mathcal{C}\partial_z\sigma_0\tau_0-i\tau_0(\partial_x\sigma_x+\partial_y\sigma_y+\partial_z\sigma_z)\nonumber\\
H_{-}(\mathbf{r})&=&i\mathcal{C}\partial_z\sigma_0\tau_0-i\tau_0(\partial_x\sigma_x+\partial_y\sigma_y-\partial_z\sigma_z)
\label{TRB-hamil}
\end{eqnarray}
The time reversal operator $\mathcal{T}=-i\tau_x\sigma_y\mathcal{K}$, commutes with the Hamiltonians in Eq.(\ref{TRS-hamil}) i.e., $[H_{\pm}(\mathbf{r}),\mathcal{T}]=0$ whereas it does not commute with the Hamiltonians in Eq.(\ref{TRB-hamil}). In addition there exists an symmetry operation
\begin{eqnarray}
\mathcal{U}=i\tau_y\sigma_y\mathcal{R}_x
\end{eqnarray}
where $\mathcal{R}_y$ is the reflection operator about the $yz$-plane. In absence of tilt, the opposite chiral sector follow the symmetry: $\mathcal{U}H_+(\mathbf{r})\mathcal{U}^{-1}=H_-(\mathbf{r})$. We call $\mathcal{U}$ as the $\mathbb{Z}_2$ (exchange) operator. The TRS tilt preserve the $\mathbb{Z}_2$ symmetry whereas TRS broken tilt breaks this symmetry. Thus the TRS broken tilt breaks both the  $\mathbb{Z}_2$ symmetry and $\mathcal{T}$ simultaneously. However, the TRS broken tilted system preserve the combined symmetry defined by the product of $\mathcal{T}$ and $\mathcal{U}$ i.e.,
\begin{eqnarray}
\mathcal{T}\mathcal{U} H_{+}(\mathbf{r}) (\mathcal{T}\mathcal{U})^{-1}=H_-(\mathbf{r})
\end{eqnarray}
The BdG Hamiltonian for the positive chirality sector is given by,
\begin{eqnarray}
\mathcal{H}^{+}_{BdG}(\phi)=\begin{pmatrix}
\mathcal{H}_{BdG}(\phi) & \emptyset\\
\emptyset & \mathcal{H}'_{BdG}(\phi)
\end{pmatrix}
\label{BdG-hamilpA}
\end{eqnarray}
in which the Hamiltonians $\mathcal{H}_{BdG}(\phi)$ is given by,
\begin{eqnarray}
\mathcal{H}_{BdG}(\phi)=\begin{pmatrix}
H_1(\mathbf{r})-\mu\sigma_0 & \Delta e^{i sgn(z)\phi/2}\\
\Delta e^{-i sgn(z)\phi/2} & \mu\sigma_0-H^*_1(\mathbf{r})
\end{pmatrix}
\end{eqnarray}
and $\mathcal{H}'_{BdG}(\phi)$ is given by,
\begin{eqnarray}
\mathcal{H}'_{BdG}(\phi)=\begin{pmatrix}
H_3(\mathbf{r})-\mu\sigma_0 & \Delta e^{i sgn(z)\phi/2}\\
\Delta e^{-i sgn(z)\phi/2} & \mu\sigma_0-H^*_3(\mathbf{r})
\end{pmatrix}
\end{eqnarray}
We define extended time reversal operator $\mathcal{T}_{BdG}$ as,
\begin{eqnarray}
\mathcal{T}_{BdG}=\begin{pmatrix}
-i\sigma_y\mathcal{K} & 0\\
0 & -i\sigma_y\mathcal{K}
\end{pmatrix}
\end{eqnarray}
By using $\mathcal{T}_{BdG}$, we find for a TRS tilt,
\begin{eqnarray}
\mathcal{T}_{BdG}\mathcal{H}_{BdG}(\phi)\mathcal{T}^{-1}_{BdG}=\mathcal{H}'_{BdG}(-\phi)
\label{sym-bdg}
\end{eqnarray} 
The BdG equation for the Hamiltonian $\mathcal{H}_{BdG}(\phi)$ is described by,
\begin{eqnarray}
\mathcal{H}_{BdG}(\phi)\psi_n=E_{n}(\phi)\psi_n
\label{bdg-eq1}
\end{eqnarray}
where $E_n$ and $\psi_n$ are eigenstate and eigenfunction labeled by an index $n$. By using Eq.(\ref{sym-bdg}), the BdG equation Eq.(\ref{bdg-eq1}) can be transformed to
\begin{eqnarray}
\mathcal{H}'_{BdG}(-\phi)\mathcal{T}_{BdG}\psi_n=E_n(\phi)\mathcal{T}_{BdG}\psi_n
\label{bdg-eq2}
\end{eqnarray}
From Eqs.(\ref{bdg-eq1}) and (\ref{bdg-eq2}), it is clear that $\mathcal{H}_{BdG}(\phi)$ and $\mathcal{H}'_{BdG}(-\phi)$ have same eigenvalues which leads to the following symmetry,
\begin{eqnarray}
E_n(\phi)=E'_n(-\phi).
\end{eqnarray}
Using the formula for the Josephson current,
\begin{eqnarray}
J(\phi)=-\frac{2e}{\hbar}\sum_{n}{\partial E_n(\phi) \over \partial \phi}f(E_n)
\end{eqnarray}
one can find the following relation for the Josephson current.
\begin{eqnarray}
J(\phi)=-J'(-\phi)
\end{eqnarray}
However, for a TRS broken tilt the Hamiltonians $\mathcal{H}_{BdG}(\phi)$ and $\mathcal{H}'_{BdG}(\phi)$ are equal i.e., $\mathcal{H}_{BdG}(\phi)=\mathcal{H}'_{BdG}(\phi)$. Consequently, the Josephson currents are equal i.e., $J(\phi)=J'(\phi)$. We now construct BdG Hamiltonian $H^-_{BdG}(\phi)$ in a similar way. It can be easily shown that,
\begin{eqnarray}
\mathcal{U}_{BdG}H^+_{BdG}(\phi)\mathcal{U}_{BdG}^{-1}=H^-_{BdG}(\phi)\nonumber\\
\label{BdG-symTRS}
\end{eqnarray}
holds for a TRS tilt with $\mathcal{U}_{BdG}=diag\{\mathcal{U},\mathcal{U}\}$. The symmetry of Eq.(\ref{BdG-symTRS}) leads to the symmetry in ABS as $E^{+}_n(\phi)=E^{-}_n(\phi)$ and consequently in the Josephson current as $J^{+}(\phi)=J^-(\phi)$. Thus the chirality Josephson current vanishes in this model system. In contrast, for a TRS broken tilt, the symmetry of BdG Hamiltonian of opposite chirality sectors is given by,
\begin{eqnarray}
{(\mathcal{T}\mathcal{U})}_{BdG}H^+_{BdG}(\phi){(\mathcal{T}\mathcal{U})}^{-1}_{BdG}=H^-_{BdG}(-\phi)
\label{BdG-symTRSB}
\end{eqnarray}
The above symmetry in Eq.(\ref{BdG-symTRSB}) leads to the symmetry in ABS as $E^{+}_n(\phi)=E^{-}_n(-\phi)$ and consequently in the Josephson current as $J^{+}(\phi)=-J^-(-\phi)$. Thus, $J^{+}(\phi)\neq J^-(\phi)$ which produces a finite chirality Josephson current in this model system.


\begin{thebibliography}{0}
\bibitem{Huang-NatCom15} S.-M. Huang, S.-Y. Xu, I. Belopolski, C.-C. Lee, G. Chang, B. Wang, N. Alidoust, G. Bian, M. Neupane, C. Zhang, S. Jia, A. Bansil, H. Lin and M. Z. Hasan, Nat. Commun. 6, 7373 (2015)
\bibitem{Xu-Sience15} S.-Y. Xu, I. Belopolski, N. Alidoust, M. Neupane, G. Bian, C. Zhang, R. Sankar, G. Chang, Z. Yuan, C.-C. Lee, S.-M. Huang, H. Zheng, J. Ma, D. S. Sanchez, B. Wang, A. Bansil, F. Chou, P. P. Shibayev, H. Lin, S. Ma, M. Z. Hasan, Science 349, 613 (2015)
\bibitem{Weng-PRX15} H. Weng, C. Fang, Z. Fang, B. A. Bernevig, and X. Dai, Phys. Rev. X 5, 011029 (2015)
\bibitem{Yang-NatPhy15} L. X. Yang, Z. K. Liu, Y. Sun, H. Peng, H. F. Yang, T. Zhang, B. Zhou, Y. Zhang, Y. F. Guo, M. Rahn, D. Prabhakaran, Z. Hussain, S.-K. Mo, C. Felser, B. Yan, and Y. L. Chen, Nat. Phys 11, 728 (2015)
\bibitem{Hirayama-PRL15} M. Hirayama, R. Okugawa, S. Ishibashi, S. Murakami, and T. Miyake, Phys. Rev. Lett. 114, 206401 (2015)
\bibitem{Ruan-Nat16} J. Ruan, S.-K. Jian, H. Zhang, S.-C. Zhang, and D. Xing, Nat. Commun. 7, 11136 (2016)
\bibitem{Burkov-PRL11} A. A. Burkov and L. Balents, Phys. Rev. Lett. 107, 127205 (2011)
\bibitem{Wan-PRB11} X. Wan, A. M. Turner, A. Vishwanath, and S. Y. Savrasov, Phys. Rev. B 83, 205101 (2011)
\bibitem{Xu-Science15} S. Y. Xu, I. Belopolski, N. Alidoust, M. Nupane, G. Bian, C. L. Zhang, R. Sankar, G. Q. Chang, Z. J. Yuan, C. C. Lee, S. M. Huang, H. Zheng, J. Ma, D. S. Sanchez, B. K. Wang, A. Bansil, F. C. Chou, P. P. Shibayev, H. Lin, S. Jia, and M. Z. Hasan, Science 349, 613 (2015)
\bibitem{Lv-PRX15} B. Q. Lv, H. M. Weng, B. B. Fu, X. P. Wang, H. Miao, J. Ma, P. Richard, X. C. Huang, L. X. Zhao, G. F. Chen, Z. Fang, X. Dai, T. Qian, and H. Ding, Phys. Rev. X 5, 031013 (2015)


\bibitem{Armitage-RMP18} N. P. Armitage, E. J. Mele, and A. Vishwanath, Rev. Mod. Phys. 90, 015001 (2018)
\bibitem{Balents-PRB12} G. B. Halasz and L. Balents, Phys. Rev. B 85, 035103 (2012)



\bibitem{Ma-Nat17} Q. Ma, S.-Y. Xu, C.-K. Chan, C.-L. Zhang, G. Chan, Y. Lin, W. Xie, T. Palacios, H. Lin, S. Jia, P. A. Lee, P. J.-Herrero, and N. Gedik, Nat. Phys 13, 842 (2017)
\bibitem{Ghosh-PRB20} S. Ghosh, D. Sinha, S. Nandy, and A. Taraphder, Phys. Rev. B 102, 121105(R) (2020)
\bibitem{Trauzettel-PRL18} S.-B. Zhang, J. Erdmenger, and B. Trauzettel, Phys. Rev. Lett 121, 226604 (2018)
\bibitem{Yang-PRL15} S. A. Yang, H. Pan, and F. Zhang, Phys. Rev. Lett 115, 156603 (2015)
\bibitem{Heidari-PRB20} S. Heidari and R. Asgari, Phys. Rev. B 101, 165309 (2020)
\bibitem{Simon-PRB19} S. Bertrand, J.-M. Parent. R. Cote, and I. Garate, Phys. Rev. B 100, 075107 (2019)
\bibitem{Simon-PRB17} S. Bertrand, I. Garate, and R. Cote, Phys. Rev. B 96, 075126 (2017)




\bibitem{Pesin-PRL17} J. F. Steiner, A. V. Andreev, and D. A. Pesin, Phys. Rev. Lett 119, 036601 (2017)
\bibitem{Burkov-PRL14} A. A. Burkov, Phys. Rev. Lett 113, 187202 (2014)
\bibitem{Uchida-Jpn14} S. Uchida, T. Habe, and Y. Asano, J. Phys. Soc. Jpn. 83, 064711 (2014)
\bibitem{Bovenzi-PRL17} N. Bovenzi, M. Breitkreiz, P. Baireuther, T. E. O'Brien, J. Tworzydlo, I. Adagideli, and C. W. J. Beenakker, Phys. Rev. B 96, 035437 (2017)
\bibitem{Zyuzin-PRB12} A. A. Zyuzin and A. A. Burkov, Phys. Rev. B 86, 115133 (2012)
\bibitem{Son-PRB13} D. T. Son and B. Z. Spivak, Phys. Rev. B 88, 104412 (2013)


\bibitem{Yu-PRL18} W. Yu, W. Pan, D. L. Medlin, M. A. Rodriguez, S. R. Lee, Z.-q. Bao, and F. Zhang, Phys. Rev. Lett 120, 177704 (2018)
\bibitem{Li-Nat18} C. Li, J. C. de Boer, B. de Ronde, S. V. Ramankutty, E. Van Heumen, Y. Huang, A. A. Golubov, M. S. Golden, and A. Brinkman, Nat. Mater. 17, 875 (2018)
\bibitem{Madsen-PRB17} K. A. Madsen, E. J. Bergholtz, P. W. Brouwer, Phys. Rev. B 95, 064511 (2017)
\bibitem{Khanna-PRB16} U. Khanna, D. K. Mukherjee, A. Kundu, and S. Rao, Phys. Rev. B 93, 121409(R)(2016)

\bibitem{Sinha-PRB20} D. Sinha, Phys. Rev. B 102, 085144 (2020)


\bibitem{Soluyanov-Nat15} A. A. Soluyanov, D. Gresch, Z. Wang, Q. S. Wu, M. Troyer, X. Dai, and B. A. Bernevig, Nature 527, 495 (2015)
\bibitem{Jiang-Nat17} J. Jiang, Z. K. Liu, H. F. Yang, C. R. Rajamathi, Y. P. Qi, L. X. Yang, C. Chen, H. Peng, C.-C. Hwang, S. Z. Sun, S.-K. Mo, I. Vobornik, J. Fujii, S. S. P. Parkin, C. Felser, B. H. Yan, and Y. L. Chen, Nat. Comm 8, 13973 (2017)
\bibitem{Wang-PRL16} Z. Wang, D. Gresch, A. A. Soluyanov, W. Xie, S. Kushwaha, X. Dai, M. Troyer, R. J. Cava, and B. A. Bernevig, Phys. Rev. Lett. 117, 056805 (2016)
\bibitem{Li-Nat17} P. Li, Y. Wen, X. He, Q. Zhang, C. Xia, Z.-M. Yu, S. A. Yang, Z. Zhu, H. N. Alshareef, and X.-X. Zhang, Nat. Commun 8, 2150 (2017)





\bibitem{Hou-PRB17} Z. Hou and Q.-F. Sun, Phys. Rev. B 96, 155305 (2017)
\bibitem{Faraei-PRB19} Z. Faraei and S. A. Jafari, Phys. Rev. B 100, 245436 (2019)
\bibitem{Faraei-PRB20} Z. Faraei and S. A. Jafari, Phys. Rev. B 101, 214508 (2020)

\bibitem{Sinha-EPJB} D. Sinha, Eur. Phys. J. B 92, 61 (2019)
\bibitem{Chan-PRB17} C.-K. Chan, N. H. Lindner, G. Refael, and P. A. Lee, Phys. Rev. B 95, 041104(R) (2017)
\bibitem{Beenakker-PRL16} T. E. O'Brien, M. Diez, and C. W. J. Beenakker, Phys. Rev. Lett. 116, 236401 (2016)
\bibitem{Trescher-PRB15} M. Trescher, B. Sbierski, P. W. Brouwer, and E. J. Bergholtz, Phys. Rev. B 91, 115135 (2015)
\bibitem{Lee-PRB15} C.-C. Lee, S.-Y. Xu, S.-M. Huang, D. S. Sanchez, I. Belopolski, G. Chang, G. Bian, N. Alidoust, H. Zheng, M. Neupane, B. Wang, A. Bansil, M Z. Hasan, and H. Lin, Phys. Rev. B 92, 235104 (2015)




\bibitem{Linder-PRB09} J. Linder, A. M. Black-Schaffer, T. Yokoyama, S. Doniach, and A. Sudb\o, Phys. Rev. B 80, 094522 (2009)
\bibitem{Kulikov-PRB20} K. Kulikov, D. Sinha, Y. M. Shukrinov, and K. Sengupta, Phys. Rev. B 101, 075110 (2020)

\bibitem{Tanaka-PRB97} Y. Tanaka and S. Kashiwaya, Phys. Rev. B 56, 892 (1997)
\bibitem{Buzdin-RMP05} A. I. Buzdin, Rev. Mod. Phys. 77, 935 (2005)
\bibitem{Buzdin-PRL08} A. Buzdin, Phys. Rev. Lett. 101, 107005 (2008)
\bibitem{Yokoyama-PRB14} T. Yokoyama, M. Eto, Y. V. Nazarov, Phys. Rev. B 89,195407 (2014)
\bibitem{Dolcini-PRB15} F. Dolcini, M. Houzet, and J. S. Meyer, Phys. Rev. B 92, 035428 (2015)
\bibitem{tanaka-PRL09} Y. Tanaka, T. Yokoyama, and N. Nagaosa, Phys. Rev. Lett 103, 107002 (2009)
\bibitem{Linder-PRL10} J. Linder, Y. Tanaka, T. Yokoyama, A. Sudb\o, and N. Nagaosa, Phys. Rev. Lett 104, 067001 (2010)





\bibitem{Samuelsson-PRB00} P. Samuelsson, J. Lantz, V. S. Shumeiko, and G. Wendin, Phys. Rev. B 62, 1319 (2000)
\bibitem{Beenakker-PRL13} C. W. J. Beenakker, D. I. Pikulin, T. Hyart, H. Schomerus, and J. P. Dahlhaus, Phys. Rev. Lett 110, 017003 (2013)
\bibitem{Crepin-PRL14} F. Crepin and B. Trauzettel, Phys. Rev. Lett 112, 077002(2014)
\bibitem{Crepin-Physica16} F. Crepin and B. Trauzettel, Physica E 75, 379 (2016)
\bibitem{Burkov-PRL18} A. A. Burkov, Phys. Rev. Lett 120, 016603 (2018)
\bibitem{Semenoff-PRL84} G. W. Semenoff, Phys. Rev. Lett 53, 2449 (1984)


\bibitem{Okugawa-PRB17} R. Okugawa and S. Murakami, Phys. Rev. B 96, 115201 (2017)
\bibitem{Kim-PRB16} H. Kim and S. Murakami, Phys. Rev. B 93, 195138 (2016)
\end{thebibliography}
\end{document}